 \documentclass{aa} 
\usepackage{graphicx}
 \usepackage[varg]{txfonts}
\usepackage[draft,colorlinks=true,citecolor=blue,linkcolor=blue]{hyperref}

\usepackage{natbib}

\usepackage[varg]{txfonts}
\usepackage{graphicx}
\usepackage{lscape}
\usepackage{array}
\usepackage{subcaption}
\graphicspath{{./figs/}}

\usepackage{multirow}
\usepackage{mathtools}
\usepackage{multirow, makecell}
\usepackage[dvipsnames]{xcolor}

\begin{document}

\title{High-resolution X-ray spectroscopy of the stellar wind in Vela X-1 during a flare}

   \author{
M.~Lomaeva\inst{1,2}\and
V.~Grinberg\inst{3} \and
M.~Guainazzi\inst{2} \and 
N. Hell\inst{4} \and
S. Bianchi\inst{5} \and
M. Bissinger n\'e K\"uhnel{\inst6} \and
F. F\"urst\inst{7}\and
P.~Kretschmar\inst{7} \and
M. Mart\'inez-Chicharro\inst{8} \and
S. Mart\'inez-N\'u\~nez\inst{9} \and
J.M. Torrej\'{o}n\inst{8}
}

\institute{Department of Physics and Astronomy, University College London, Gower Street, London WC1E 6BT, UK \\
    \texttt{Maria.Lomaeva.19@ucl.ac.uk}
    \and ESA European Space Research and Technology Centre (ESTEC), Keplerlaan 1, 2201 AZ, Noordwijk, The Netherlands\label{1} 
    \and Institut f\"ur Astronomie und Astrophysik, Eberhard Karls Universität T\"ubingen, Sand 1, 72076 T\"ubingen, Germany
    \and Lawrence Livermore National Laboratory, 7000 East Avenue, Livermore, CA 94550, USA
    \and Dipartimento di Matematica e Fisica, Universit\`a degli Studi Roma Tre, via della Vasca Navale 84, 00146 Roma, Italy
     \and Dr. Remeis Sternwarte \& ECAP, Universit\"at Erlangen-N\"urnberg, Sternwartstr. 7, 96049 Bamberg, Germany
    \and 
European Space Astronomy Centre (ESAC), Science Operations Department,
 E-28692, Villanueva de la Ca\~{n}ada, Madrid, Spain
 \and Instituto Universitario de Física Aplicada a las Ciencias y las Tecnologías, Universiad de Alicante, Alicante E-0380, Spain
    \and
Instituto de F\'isica de Cantabria (CSIC-Universidad de Cantabria), E-39005, Santander, Spain    
}

\date{ -- / --}

\abstract
    {We present a $\sim$130~ks observation of the prototypical wind-accreting, high-mass X-ray binary Vela X-1 collected with \textit{XMM-Newton} at orbital phases between 0.12 and 0.28. A strong flare took place during the observation that allows us to investigate the reaction of the clumpy stellar wind to the increased X-ray irradiation.}
    {To examine the wind's reaction to the flare, we performed both time-averaged and time-resolved analyses of the RGS spectrum and examined potential spectral changes.}
    {We focused on the high-resolution \textit{XMM-Newton} RGS spectra and divided the observation into pre-flare, flare, and post-flare phases. We modeled the time-averaged and time-resolved spectra with phenomenological components and with the self-consistent photoionization models calculated via {\tt CLOUDY} and {\tt XSTAR} in the pre-flare phase, where strong emission lines due to resonant transitions of highly ionized ions are seen. }
    {In the spectra, we find emission lines corresponding to K-shell transitions in highly charged ions of oxygen, neon, magnesium, and silicon as well as radiative recombination continua (RRC) of oxygen. Additionally, we observe potential absorption lines of magnesium at a lower ionization stage and features identified as iron L lines. The \texttt{CLOUDY} and \texttt{XSTAR} photoionization models provide contradictory results, either pointing towards uncertainties in theory or possibly a more complex multi-phase plasma, or both.}
    {We are able to demonstrate the existence of a plethora of variable narrow features, including the firm detection of oxygen lines and RRC that RGS enables to observe in this source for the first time. We show that Vela X-1 is an ideal source for future high-resolution missions, such as \textit{XRISM} and \textit{Athena}.} 


\maketitle
\section{Introduction}\label{sec:intro}

\subsection{Vela X-1}\label{sec:intro:velax1}
\object{Vela X-1} (4U\,0900$-$40) is an eclipsing high mass X-ray binary (HMXB) in which a neutron star (NS)
 closely orbits a supergiant mass donor with a period of about nine days \citep{hiltner_72,forman_73}, accreting from the strong stellar wind and emitting brightly
in the X-rays. The system lies relatively close, with a Gaia distance estimate of 
$2.42^{+0.19}_{-0.17}$~kpc \citep{Bailer-Jones:2018}, such that despite its intrinsically moderate
X-ray luminosity \citep[$\sim$4$\times$10$^{36}$~erg~s$^{-1}$,][]{Nagase:86}, it is among the
brightest persistent sources on the X-ray sky and has been well-observed ever since its
detection \citep{Chodil:67}. It is often considered as a ``prototype'' for wind-accreting HMXB, and
its system parameters or approximations have been used in various modeling studies 
 \citep[see, e.g.,][]{Blondin:90,Blondin:91, ManousakisWalterBlondin:2014PMB,elmellah_2019}.

The mass donor, HD~77581, has a mass of $\sim$24\,$M_\odot$ and a radius of about 30\,$R_\odot$ 
\citep{kerkwijk_95,quaintrell_03,Rawls:2011}. Its mass loss rate has been estimated by different authors 
over the years -- see \citet{Kretschmar:2019INT} for an overview --
to be in the range of $\sim$0.5--2$\times$10$^{-6} M_\odot$ yr$^{-1}$, 
while estimates for the terminal velocity of the wind, $v_\infty$, have ranged
from 1700\,km\,s$^{-1}$ \citep{Dupree:80} to values of
600--700\,km\,s$^{-1}$ in more recent studies \citep{krticka_12, gimenez-garcia_16,sander_18}.

The X-ray source is a relatively massive NS, with estimates ranging from $\sim$1.8\,$M_\odot$  to $\sim$2.3\,$M_\odot$ \citep{Barziv:2001,quaintrell_03,Rawls:2011}. It orbits HD77581 at an average distance of 53.4 R$_\odot$ or 1.7 R$_\textnormal{HD 77581}$ on an orbit with low  eccentricity \citep[$e=0.0898\pm0.0012$,][]{Bildsten:97}. It is a persistent X-ray pulsar with a pulse period of $\sim$283~s, which varies randomly on longer timescales \citep{Deeter:87AJ,Bildsten:97}. The pulsar has a high magnetic field of a few $10^{12}$\,G that can be directly measured from the detected cyclotron lines \citep[e.g.,][and references therein]{Fuerst_2014a}. Being deeply embedded in the dense stellar wind close to the mass donor, the neutron star shows  highly variable accretion and absorption depending on the changing properties and structure of the stellar wind \citep[e.g.,][]{Malacaria_2016a,Fuerst:2010}, but can also influence the wind flow through its X-ray radiation \citep[e.g.,][]{Blondin:90,krticka_12,sander_18}.

The eclipse of the X-ray source in \object{Vela X-1} covers $\sim$20\% of the orbit duration, \citep[e.g.,][]{Falanga:2015}. In the literature, orbital phases are sometimes expressed relative to the mid-eclipse time, sometimes relative to the time of mean longitude ($T_{90}$ or $T_{\pi/2}$ used in this work). \citet{Kreykenbohm_08} determined  the offset between these two definitions as 0.026$\pm$0.005 in orbital phase.

After eclipse egress, the observed absorption (characterized by the equivalent hydrogen column density, $N_\textnormal{H}$) usually decreases strongly by more than an order of magnitude, as the neutron star slowly emerges from behind the extended atmosphere or inner wind base of the supergiant star \citep{Nagase:86,Lewis:92}, often reaching minimal absorption around orbital phase 0.25, where the neutron star is mainly moving towards the observer \citep{Doroshenko_2013a}. At later phases, the observed $N_\textnormal{H}$ can vary strongly by an order of magnitude \citep[e.g.,][]{Orlandini_1998a} and shows different peaks and minima in different observations, but always also a marked overall increase towards eclipse ingress \citep[e.g.,][]{Nagase:86,HaberlWhite:90,Malacaria_2016a}.  The evolution of $N_\textnormal{H}$  is usually attributed to the changing sight-line through the complex accretion  geometry, which may include an accretion wake, a photoionization wake, and a tidal stream (e.g., \citealt{Blondin:90,Kaper_1994a,van_Loon_2001a};  see \citealt{Grinberg_2017a} for a visualisation and further references).  On top of the changes along the orbit, there is significant irregular variation both in absorption as in intrinsic X-ray flux, including bright flares and states of very low flux known as ``off states'' \citep[e.g.,][]{HaberlWhite:90,Kreykenbohm_08,Odaka_2013a,Martinez-Nunez_2014a,Sidoli_2015a}.

Early X-ray line studies in \object{Vela X-1} have mainly been focused on the prominent iron line complex, see \citet[][\textit{OSO 8}]{Becker:78} and \citet[][\textit{Tenma}]{ohashi_1984}, for example. \citet[][\textit{Tenma}]{Sato:86} also detected a broad Fe K$\alpha$ emission line, but they suspected that fluorescent K$\alpha$ lines of Si, S, Ar, Ca, and Ni, as well as the Fe~K$\beta$ line also contribute to the overall line intensity. Various further elements have been reported from spectra taken during, or close to eclipse \citep{nagase_94,sako_99,schulz_02}, including recombination lines and radiative recombination continua (RRC), as well as fluorescence lines. The variety of ionization states strongly indicates the presence 
of an inhomogeneous wind with optically thick, less ionized matter coexisting with warm photoionized plasma, for example, in the form of clumps intrinsic to the wind or in the accretion wake \citep{Grinberg_2017a}.

\citet{Goldstein_2004a} and \citet{watanabe_06} each analyzed \textit{Chandra}-HETGS data from three orbital phases: in eclipse, around phase
0.25, and around phase 0.5. In eclipse and in the heavily absorbed observation at phase 0.5, a large number
of emission lines were identified, while around phase 0.25, the spectrum was dominated by the continuum and the line content was not discussed further.
\citet{Grinberg_2017a} revisited the \textit{Chandra}-HETGS observation around phase 0.25, creating separate spectra for time intervals of low and high spectral hardness, that is, performing an absorption-resolved analysis. They detected line features from high and low 
ionization species of Si, Mg, and Ne, as well as strongly variable absorption. Again, this implies 
co-existence of cool and hot gas phases that can either be explained by an intrinsically clumpy stellar wind or a highly variable, structured accretion flow close to the compact object or a combination of both effects.

\subsection{X-ray studies of winds in HMXBs and modeling efforts}

Studying stellar winds in HMXBs is not only interesting for research on accretion in X-ray binaries, they are also crucial for understanding the stellar winds from massive stars. The diagnostics of the stellar wind structure and properties obtained from X-ray observations in HMXBs are complementary to diagnostics at other wavelengths and for single stars and may be used to break degeneracies in derived models of  stellar winds from massive stars in general. \citet{Martinez-Nunez:2017} provide an extensive review of theory and observations of inhomogeneous winds in HMXBs as well as in isolated massive stars.  As described in this review, various authors have used a wide range of approaches to estimate wind parameters and indications for large and small-scale structures. These include studies of absorption variations by intervening material \citep[e.g.,][]{Grinberg_2015a,Miskovicova_2016a}, the variability of the intrinsic X-ray continuum as observed in flares and off-states \citep[e.g.,][]{Bozzo_2013a,Martinez-Nunez_2014a,Pradhan_2019a}, or fluorescence line studies \citep[e.g.,][]{Boroson_2003a,wojdowski_03}. 
It can be difficult, though, to directly compare results of particular studies, since different effects on the actual observables may be hard to disentangle and, in addition, underlying assumptions frequently vary between studies.

The huge range in length scales to be treated -- from binary system scales of $>$10$^8$\,km to accretion column scales of $<$1\,km -- as well as the many, interconnected  physical effects to be taken into account, ideally dynamically self-consistent, have so far precluded a detailed, fully realistic model of a HXMB accreting X-ray pulsar. Existing model efforts for HMXBs treat only certain aspects of the system in detail,  while others are prohibited or largely simplified. This limits the ability to test observations directly against predictions, but some comparisons are both possible and have pushed the field forward \citep[e.g.,][]{Manousakis_2015a,Grinberg_2017a}.

Previous studies of clumpy winds in HMXBs with RGS include 4U1700-37 by \cite{meer_2004} and 4U0114+65 by \cite{sanjurjo_2017}. The former work did not manage to obtain much information about the highly absorbed source from the RGS spectra due to a low count rate despite strong flares. \cite{sanjurjo_2017}, on the other hand, were able to detect several emission lines with RGS below 1 keV: \ion{C}{v} Ly$\alpha$, \ion{N}{vi} He-like triplet,  \ion{O}{vii} Ly$\alpha$,  \ion{Fe}{xvii}, and  \ion{Fe}{xviii}. No flares were detected in this observation.

In the following, we use \textsl{XMM}-RGS observations of Vela X-1 at orbital phase $\sim$0.25 during a strong flare in order to probe the stellar wind and accretion structure properties in the system and, in particular, the wind's response to changes in X-ray irradiation. We note, in particular, that this is the first analysis of Vela X-1 with RGS, enabling the coverage of lower energies than previous studies with \textsl{Chandra}-HETGS. This work makes use of previous results of the \textit{EPIC}-pn analysis of the same observation in  \cite{Martinez-Nunez_2014a}, which is briefly introduced,  together with the general discussion of the data used, in Sect.~\ref{sect:data}. We present our spectral analysis, including a blind line search and photoionization modeling in Sect.~\ref{sect:analysis}, discuss our findings in Sect.~\ref{sect:discussion}, along with an outlook on the future prospects of observations with \textsl{XRISM} and \textsl{Athena}. Finally, we present our summary in Sect.~\ref{sect:summary}.

\section{Data}
\label{sect:data}

\subsection{Observations}

Vela X-1 was observed on May 25$-$26, 2006 (MJD 53 880.439 $-$ MJD 53 881.877; \texttt{ObsID} 0406430201) using the European Photon Imaging Camera (EPIC) and the Reflection Grating Spectrometers (RGS, \citealt{herder_01}) on-board the \textit{XMM-Newton} observatory \citep{jansen_01}
during its complete revolution 1183. The observations lasted for $\sim$130 ks at orbital phases $\phi_\mathrm{orb} = 0.12$--0.28, according to the ephemeris in \cite{Kreykenbohm_08}, with phase zero defined as $\phi_\mathrm{90} = 0$. During observations, the EPIC-Metal Oxide Semi-conductor (EPIC-MOS1, \citealt{turner_01}) and RGS~1 cameras were disabled to guarantee the maximum telemetry for the \textit{EPIC}-pn camera \citep{struder_01}. However, on May 26, the RGS~1 camera was switched on since the telemetry bandwidth would still make it useful (Fig.~\ref{fig:rgs_lc}). The \textit{EPIC}-pn data from this observation were studied in \cite{Martinez-Nunez_2014a}.

\subsection{Previous analysis of simultaneous \textit{EPIC}-pn data}\label{sec:epic_pn}

Our analysis builds on the previous results obtained with \textsl{EPIC}-pn, in particular, on the knowledge won about the evolution of the continuum emission during the observation. In this section, we give a short summary of those key findings in \cite{Martinez-Nunez_2014a} that drive our analysis and interpretation.

The authors subdivided the observation into individual spectra, each of 1.1\,ks length to trace the evolution of the continuum and absorption throughout the flare. They saw a strong flare during the observation, with an increase in observed \textsl{EPIC}-pn count{\-}rate by a factor of $\gtrsim$10. They found that this increase is driven both by  an increase in the intrinsic luminosity of the source and a decrease in absorption.
The phenomenological analysis of the unabsorbed flux and hydrogen column density evolution throughout the flare suggested two physical sources were visible: the neutron star emission and the scattered emission by the wind. The authors propose that the origin of the flare might be associated with accretion of a dense wind clump onto the neutron star.

\cite{Martinez-Nunez_2014a} also noted a decrease in the column density in two of the absorbing components that they interpret as being located in the vicinity of the neutron star. This was partially explained by accretion of the dense clump and subsequent reduction of the amount of matter in the vicinity of the neutron star. Another reason given was the decrease in the amount of stellar wind material along the line of sight with orbital phase as the neutron star moves along the orbit after eclipse egress.

The \textsl{EPIC}-pn camera covers a wider energy range ($\sim$0.5--10\,keV) than the RGS camera ($\sim$0.33--2.5\,keV), however, with a significantly lower resolving power. The \textit{EPIC}-pn spectra in \cite{Martinez-Nunez_2014a} show hints of spectral line features at energies below $\sim$2 keV, and in fact, those have also been observed and identified in \textsl{Chandra}-HETG observations of Vela X-1 \citep{Goldstein_2004a,watanabe_06,Grinberg_2017a}, although none of these works addressed strong flares as seen in our observation. In this work, we aim to characterize these narrow features and their changes throughout the flare using RGS with its higher resolving power.

\subsection{Data reduction}

\begin{figure}[htbp]
\centering
\includegraphics[width=0.5\textwidth]{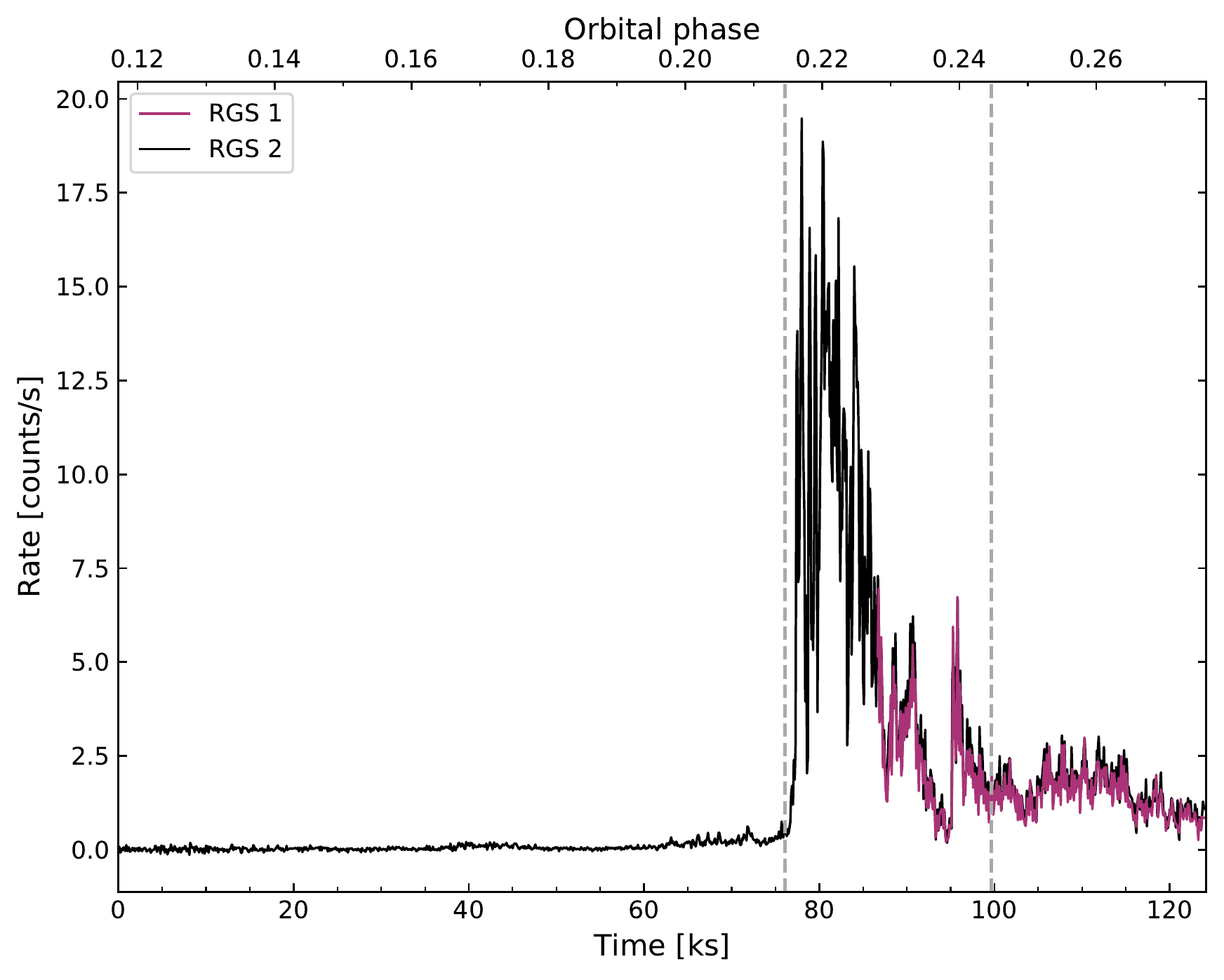}   
\caption{RGS light curves of Vela~X-1 during the observation discussed in this paper. Time on the
abscissa are from the beginning of the observations (MJD 53 880.439). 
The dashed lines show the pre-, during and post-flare phases. We use the time of mean longitude of 90 degrees, $T_{90}$, from \cite{Kreykenbohm_08} as phase zero. } 
\label{fig:rgs_lc}
\end{figure}

Data reduction was performed in \textit{XMM-Newton} Science Analysis System (SAS) v. 17.0.0 \citep{gabriel_04} using the most up-to-date calibration files available in October 2018. We processed the data with the standard SAS tool \texttt{rgsproc} to generate event files, spectra and response matrices.  We also produced exposure-corrected and background-subtracted RGS~1 and RGS~2 light curves with the  \texttt{rgslccorr} task. From that, we defined the good time intervals (GTIs) and generated three sets of spectra with the \texttt{tabgtigen} task (pre-flare, flare and post-flare phases as shown in Fig. \ref{fig:rgs_lc}). The flare phase begins after 76.1 ks ($\phi_\mathrm{orb}$ = 0.215), and the post-flare phase starts after 99.6 ks ($\phi_\mathrm{orb}$ = 0.245). 
The spectra were then rebinned with an \texttt{ftools} command  \texttt{ftgrouppha} with an optimal binning scheme from \cite{kaastra_16}. 

 The publicly available Observation Data Files (ODFs) suffer from time jumps resulting in an error message when running the \texttt{rgslccorr} task as it cannot handle these jumps properly. The correct version of the ODFs without the time jumps was received directly from the \textit{XMM-Newton} Science Operation Centre. In addition, due to failures of individual RGS CCDs\footnote{\url{https://xmm-tools.cosmos.esa.int/external/xmm_user_support/documentation/uhb/rgs.html}}, there are gaps in the spectra at $\sim$0.9--1.2\,keV for RGS~1, and $\sim$0.5--0.6\,keV for RGS~2.

As the background rate for both RGS cameras showed that the maximum recommended value of 2 counts/s was not exceeded, we did not remove intervals of high particle background. This explains why our total lightcurves (MJD 53~880.439 to MJD 53~881.878) are longer than the ones presented in \citet{Martinez-Nunez_2014a} (MJD 53~880.613 to MJD 53~881.768), who applied such a correction.

\section{Spectral analysis and results}
\label{sect:analysis}

The main goal of the spectroscopic analysis discussed in this paper is the diagnostics of the plasma in the stellar wind. For this purpose, a robust method is needed to identify and characterize narrow-band features, that is, emission and absorption lines and radiative recombination continua (RRC). We followed a two-tier approach. We first selected a list of narrow-band feature candidates through a blind search, following the method discussed in \citet{tombesi10}. The candidate line list was later used as an input to a formal forward-folding fitting procedure. For the latter step, we used \texttt{XSPEC} \citep{arnaud_96} and, in particular, \texttt{PyXspec}. Throughout this work, we use the Minuit2 \texttt{migrad} minimization method and, due to the low numbers of counts per spectral bin, the C-statistic \citep{cash_79}.
We have made use of the following data: 
1) in the pre-flare phase, only the RGS~2 spectrum was analyzed as no RGS~1 data are available;
2) in the flare phase, the spectra of RGS~1 and RGS~2 were analyzed separately and assuming independent spectral parameters for the two cameras since RGS~1  covers a much shorter part of the flare than RGS~2;
3) in the post-flare phase, RGS~1 and RGS~2 spectra were analyzed simultaneously, assuming the same spectral parameters since RGS~1 and RGS~2 cover the same time interval; cf Fig.~\ref{fig:rgs_lc}.  

To take into account possible differences in cross-calibration of the two cameras, we included a cross-normalization constant, the best-fitting parameter of which is consistent with 1 ($0.98 \pm 0.04$), as expected \citep{vries_15}.

In the remainder of this section, we introduce the continuum model in Sect.~\ref{sect:continuum}, explain the method for the blind line search in Sect.~\ref{sect:blind}, present the detected narrow-band spectral features in Sect.~\ref{sect:lines},
the direct fits of the $R$ and $G$ parameters in Sect.~\ref{sect:rg}, 
and conclude with photoinization modeling of the data in Sect.~\ref{sect:pimodels}.

\subsection{Continuum model}
\label{sect:continuum}

In the analysis, we implement a continuum model that builds on the previous model of \citet{Martinez-Nunez_2014a} and provides an appropriate fit for the spectra throughout the whole observation, that is, in the pre-, during, and post-flare phases. Our overall model, including the narrow-band components, is described by:

\begin{equation}\label{eq:model}
\begin{split}
    F(E) =   \texttt{tbabs}_1 \times \texttt{cabs} \times \texttt{powerlaw}_1 + \texttt{tbabs}_2 \times \texttt{powerlaw}_2 \\ + \sum _{i = 1} ^n  G_i + \sum _{i = 1} ^m  RRC_i\, ,
\end{split}
\end{equation}
that is, two absorbed power laws (defined by normalization and the photon index, $\Gamma$) with an addition of $n$ Gaussian absorption or emission line profiles (defined by line energy, width and flux) and $m$ RRC (\texttt{redge} in \texttt{xspec}, defined by threshold energy, plasma temperature and overall normalization).

Unlike \citet{Martinez-Nunez_2014a}, who used \textit{XMM-Newton} \textit{EPIC}-pn spectra, our model includes only two power law continuum components instead of three. Their third component mainly contributes above $\sim$2\,keV and is thus not required here given the spectral coverage of RGS. The lack of coverage at higher energies and thus of a good constraint on the underlying, non-absorbed continuum, in general prevents a direct comparison between the parameters of the continuum components in  \cite{Martinez-Nunez_2014a} and in this paper.  However, the choice of the continuum modeling does not have a significant impact on the main results discussed here: the continua in Eq.~\ref{eq:model} only represent an empirical description, and are not used to draw any physical conclusions about the state of the plasma in the stellar wind that is based on the narrow-band features that only require a good local continuum description \citep[e.g.,][]{Goldstein_2004a,Grinberg_2017a,van_den_Eijnden_2019a}.

The two power law models of our continuum have distinct normalizations and photon indices. To account for the effects occurring in the material with a given column density, $N_\textnormal{H}$, that is situated between  the X-ray source and the detector, the power law component is modified by absorption, modeled by two different contributions. We used  {\tt tbabs} \citep{wilms00} with abundances from \citet{wilms00}\footnote{Note that modern version of {\tt tbabs} automatically uses the cross-sections of \citet{Verner_1996a}.} to model the overall absorption and follow \citet{Martinez-Nunez_2014a} in further using optically-thin non-relativistic Compton scattering, {\tt cabs}. 
In the highly time-resolved analysis of \citet{Martinez-Nunez_2014a}, where short-term strong increases of absorption, especially at the very beginning of the observation, could be resolved, the inclusion of the model was crucial. Since RGS spectra cannot be analyzed on timescales of a few ks, we do not expect \texttt{cabs}  to have a significant contribution as it should only become significant at average column densities of $\gtrsim$10$^{24}$ cm$^{-2}$. 
Nevertheless, since this component was present in the original continuum model in \citep{Martinez-Nunez_2014a}, we keep it for historical and consistency reasons.
The values of the continuum parameters we obtain are summarized in Tab.~\ref{tab:rgs_model}.

\begin{table*}[htbp]
\centering
\caption{Continuum model parameters and corresponding uncertainties. The error bars were calculated for a 90\% confidence interval and two interesting parameters.
}\label{tab:rgs_model}
\footnotesize
\renewcommand{\arraystretch}{2}
\begin{tabular}{l c c c c c c c}
\hline\hline 
Phase  & $N_\textnormal{H,1}$  & $\Gamma_1$ &N$_1$  &$N_\textnormal{H,2}$ &$\Gamma_2$ &N$_2$   \\ 
  & $[10^{22}$ cm$^{-2}]$  &   &  [ph/keV/cm$^{2}$/s]&$[10^{22}$ cm$^{-2}]$ &    & [ph/keV/cm$^{2}$/s]\\\hline 
\multirow{1}{1.2 cm}{Time-averaged}& 0.52 $^{+0.04}_{-0.22}$   & $-0.4^{+1.8}_{-1.6}$   &  (1.2 $^{+0.1}_{-0.6})\times10^{-2}$   &  2.54 $^{+0.05}_{-0.15}$   &  1.95 $^{+0.07}_{-0.24}$ &  0.16$^{+0.01}_{-0.05}$& RGS~2\\ \hline 
Pre-flare & $-$  & $-$  & $-$  & $< 0.09$ &  -1.7 $^{+0.3}_{-0.2}$  & (4.9 $^{+0.9}_{-0.4}) \times 10^{-4}$  & RGS~2\\ \hline
\multirow{2}{0.7 cm}{Flare} & 3.3 $^{+0.1}_{-0.6}$  &  2.6 $^{+0.1}_{-0.7}$  & 0.80 $^{+0.07}_{-0.34}$ & 0.9 $^{+0.4}_{-0.3}$ & 1.2 $^{+1.5}_{-2.4}$  & (4.2 $^{+6.9}_{-2.2}) \times 10^{-2}$& RGS~1\\
& 0.9  $^{+0.2}_{-1.1}$ & 1.0 $^{+0.7}_{-0.5}$ & 0.11 $^{+0.05}_{-0.02}$ & 3.27 $^{+0.06}_{-0.43}$ & 2.5 $^{+1.2}_{-0.5}$ & 1.26 $^{+0.08}_{-0.44} $&  RGS~2 \\ \hline 
Post-flare  & 0.9 $^{+1.1}_{-0.2}$ & $-1.1^{+4.0}_{-1.3}$ & (8.9$^{+11.8}_{-4.5} ) \times 10^{-3}$ & $2.1^{+5.0}_{-0.2}$ & 1.8 $^{+5.7}_{-1.4}$ &  0.2  $^{+0.2}_{-0.2} $&  RGS~1 \& RGS~2

\\ \hline
\end{tabular}
\tablefoot{ $N_\textnormal{H,1}$ refers to the hydrogen column density of both  \texttt{cabs} and \texttt{tbabs}$_1$. The cross-normalization constant used for the simultaneous fit of the post-flare RGS~1 and RGS~2 spectra is $0.98^{+0.04}_{-0.04}$.}

\end{table*} 

\begin{figure}
    \centering
    \includegraphics[width=0.5\textwidth]{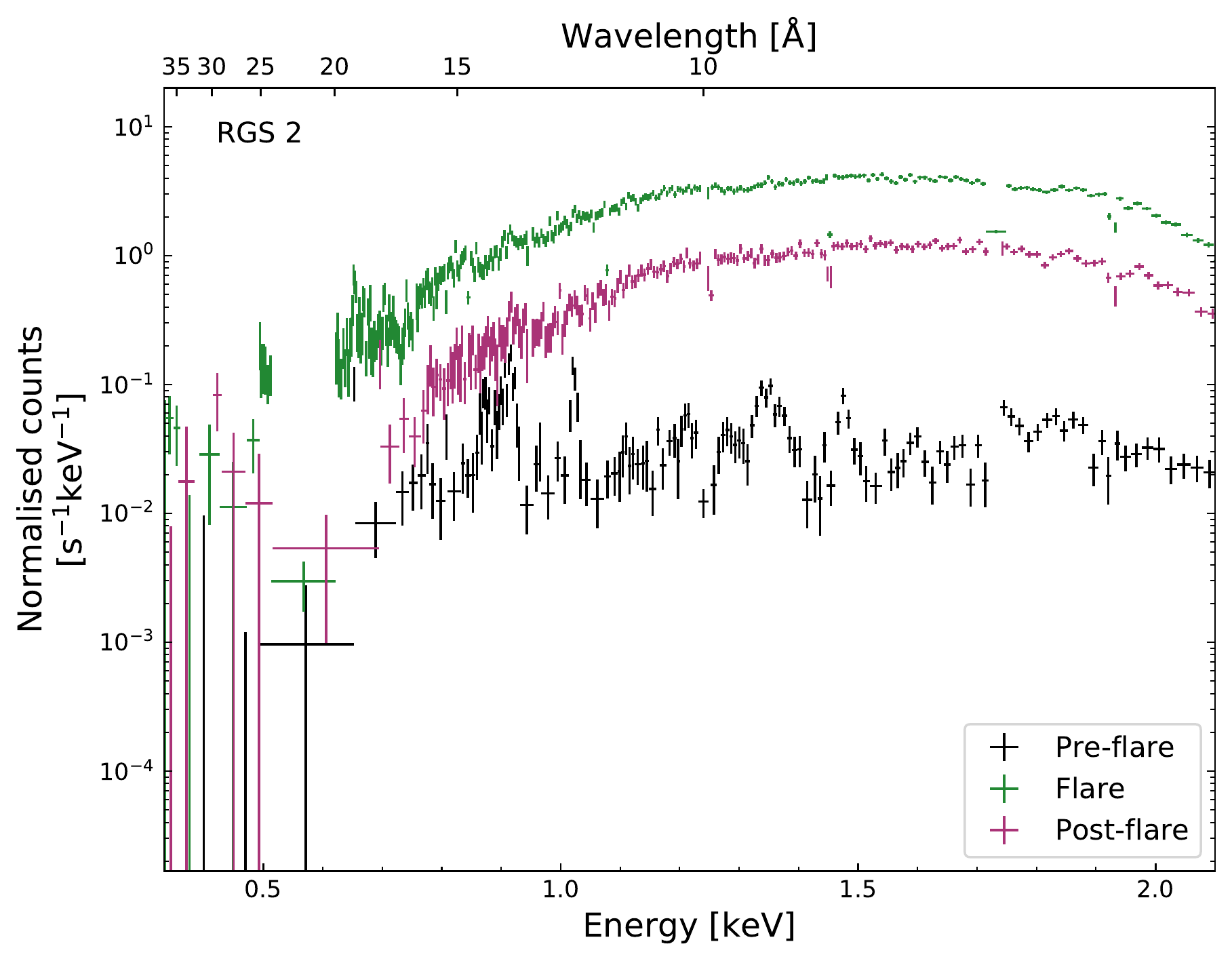}
    \caption{Re-binned, folded, and time-resolved RGS~2  count spectra (crosses).}
    \label{fig:spectra_all_phases}
\end{figure}

In Fig.~\ref{fig:spectra_all_phases}, we compare the broadband
instrument-response corrected RGS~2 spectra during the three phases. 
The quiet pre-flare phase shows a flat, very low continuum contribution with many strong narrow emission features visible already by eye. The flare causes the continuum to increase drastically as observed by both RGS cameras, and its contribution remains high compared to the pre-flare phase, even in the post-flare phase. \citet{Martinez-Nunez_2014a} saw a similar behavior in their analysis of the simultaneous \textit{EPIC}-pn spectra. In particular, they could show that the difference between the observed pre- and post-flare continuum is driven both by changes in the intrinsic flux of the power law and by changes in absorption, while  the underlying spectral shape remained fairly stable.


\subsection{Blind line search}
\label{sect:blind}

To search for faint and narrow features in the time-averaged\footnote{ Integrated over the entire observation.}  and time-resolved spectra, we opted for the algorithm introduced by \cite{tombesi10} which generates energy-intensity contour plots. The blind line search allows to detect spectral features that otherwise might be too weak to be detected with the naked eye.

In this approach, a Gaussian profile with centroid energy, $E$, and intensity, $I$, is added to the best-fit continuum (cf. Eq.~\ref{eq:model}) and moved through the given energy range in steps to search for spectral line-like features. We used four different spectral regions corresponding to parts of the spectrum where resonant transitions of highly ionized species of a given element are expected.  The corresponding energy ranges and numbers of steps in energy are listed in Tab.~\ref{tab:bls_interv}. Since we are interested in narrow features, a piece-wise approach to spectral analysis is appropriate and has been often used in the past in other analyses \citep[e.g.][]{Grinberg_2017a,van_den_Eijnden_2019a}. Our steps in energy correspond to $\sim$1\,eV\footnote{The sensitivity of the RGS instruments allows to resolve two lines separated by  $\Delta E \sim$1--2 eV at $\sim$0.4--0.6 keV and $\Delta E \sim$7--8\,eV at $\sim$1.4\,keV (\url{https://xmm-tools.cosmos.esa.int/external/xmm_user_support/documentation/uhb/rgsresolve.html}).}. The intensity of the line was varied  in 25 steps in the range of $-10^{-4}$ to $10^{-4}$\,ph\,s$^{-1}$\,cm$^{-2}$, thus covering both emission and absorption lines. 
The difference in goodness-of-fit statistics between the models with and without the Gaussian component were registered at each grid, allowing to build a goodness-of-fit parameter map over the energy vs. intensity plane corresponding to the 68\%, 90\% and 99\% confidence levels for one interesting parameter\footnote{See \cite{lampton_76} for definition of interesting parameters. Generally, the number of interesting parameters affects the range of their confidence intervals.} (see Fig. 3 in \cite{tombesi10} for illustration). From this map, candidate lines were drawn as those corresponding to a confidence level larger than 99\% for two interesting parameters.

We then identified the features and modeled them with Gaussian components in case of emission and absorption lines and the \texttt{redge} model in case of RRC. Several weak spectral features suggested by the blind line search were subsequently removed from our final model as they were judged to be instrumental or otherwise artificial. In particular, we disregarded a possible line at 732.01\,eV detected in a feature-rich region: the region around $\sim$730\,eV is complex, with two of our continuum components intersecting and \ion{O}{vii} RRC present \citep[at 0.739\,keV;][]{Lotz_68}. The apparent emission feature at 515.20\,eV in the RGS~1 flare spectrum on the other hand, is coincident with a cool pixel at 514.43\,eV\footnote{\url{https://xmm-tools.cosmos.esa.int/external/xmm_user_support/documentation/uhb/rgsmultipoint.html}}.

\begin{table}
\centering
\caption{Regions used for line searches and number of steps used in each region.}\label{tab:bls_interv}
\begin{tabular}{l c c c c}
\hline
\hline
Main expected element & Energy range [eV]& $N_\textnormal{steps}$  \\
\hline
O & 300--900 & 667 \\
Ne & 800--1400& 667 \\
Mg & 1200--1700 & 556 \\
Si & 1600--2100& 667 \\
\hline
\end{tabular}
\end{table}

\subsection{Narrow-band spectral features}
\label{sect:lines}

\subsubsection{N, O, Ne, Mg, and Si lines} \label{sect:lines_analysis}
\label{sect:lines:features}

All Gaussian components in our model were assumed to be unresolved (i.e., width~=~$0$) since leaving the intrinsic width free did not yield an improvement in the quality of the fit, at least at the 90\% confidence level for two interesting parameters in the time-averaged RGS~2 spectrum.
This is consistent with previous results at different orbital phases with \textsl{Chandra}-HETG where most lines were unresolved even at HETGs higher resolution \citep{Goldstein_2004a,Grinberg_2017a}. Likewise, we verified that the centroid energy of the lines with an unambiguous identification are consistent with the laboratory energy of the corresponding atomic transition. The calculated line velocities appeared to be consistent with zero which is an expected behavior given our expectation of velocities of a few hundred km/s \citep[e.g.,][]{Grinberg_2017a} and the resolution of RGS. These measurements were therefore not included in Tables~\ref{tab:lines} and~\ref{tab:all_lines}.

\begin{figure*}[htbp]
  \begin{subfigure}[b]{0.5\linewidth}\label{fig:spectra_preflare}
   \includegraphics[width=1\textwidth]{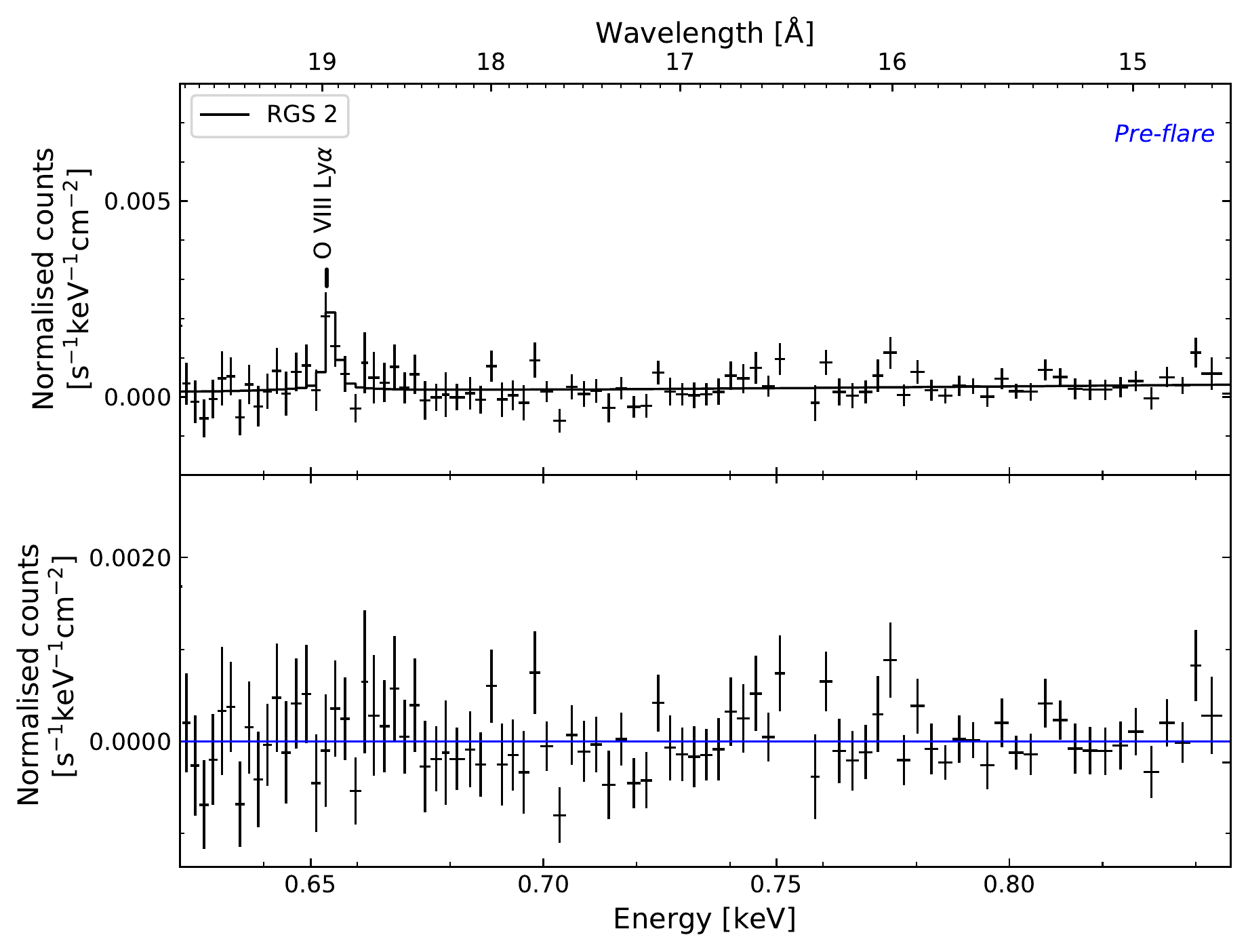} 
  \end{subfigure} 
  \begin{subfigure}[b]{0.5\linewidth}
    \includegraphics[width=\textwidth]{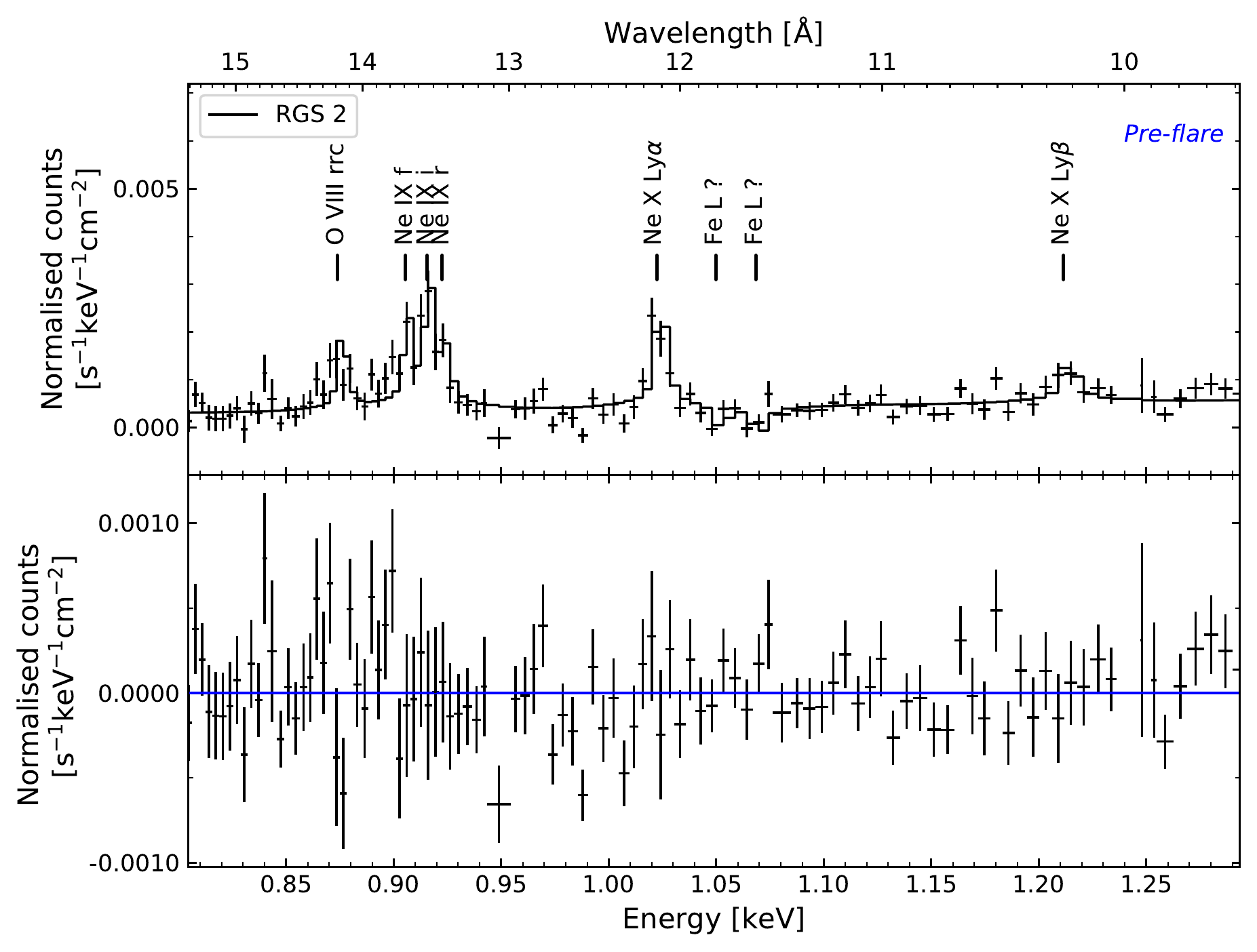}
  \end{subfigure} 
  \begin{subfigure}[b]{0.5\linewidth}
    \includegraphics[width=\textwidth]{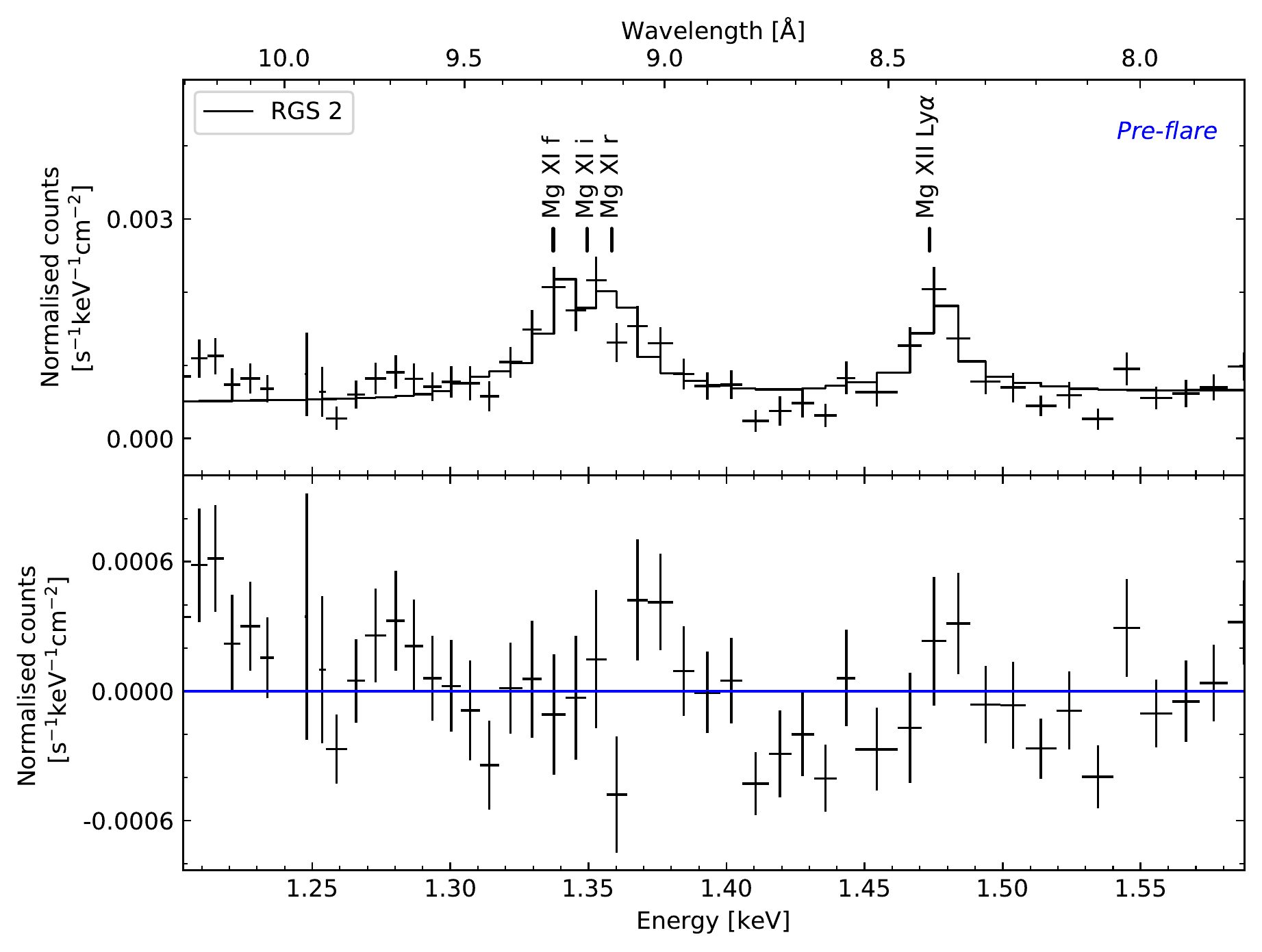}
  \end{subfigure}
  \hfill
  \begin{subfigure}[b]{0.5\linewidth}
     \includegraphics[width=\textwidth]{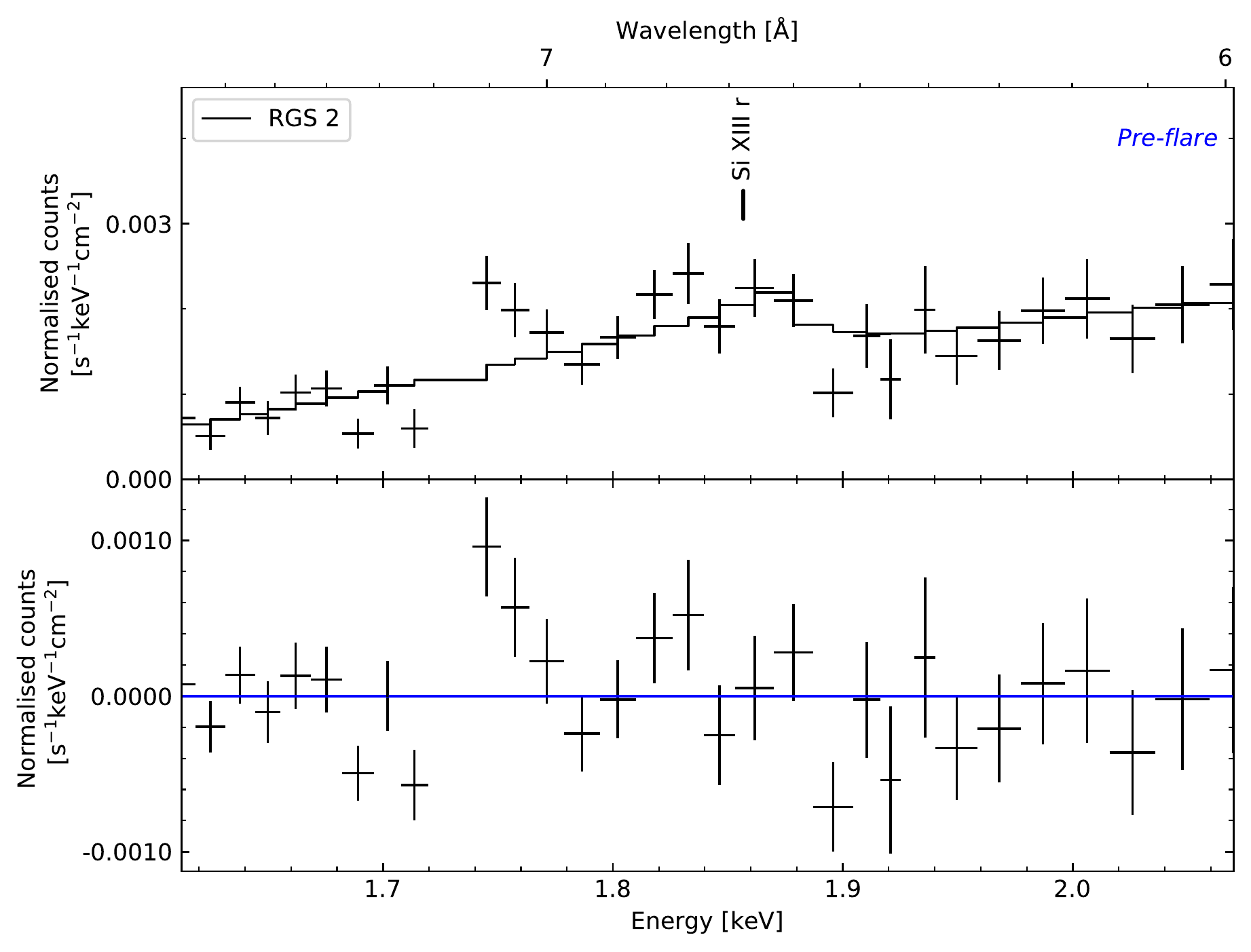}
  \end{subfigure} 
\caption{{\it Upper panel}: RGS~2 spectrum of the pre-flare phase in four energy bands ({\it crosses}), and best-fit model containing Gaussian components and RRS ({\it solid line}). {\it Lower panel}: residuals in units of data-model. } \label{fig:preflare_spectra}
\end{figure*}

\begin{figure*}[htbp]
  \begin{subfigure}[b]{0.5\linewidth}\label{fig:spectra_flare}
   \includegraphics[width=1\textwidth]{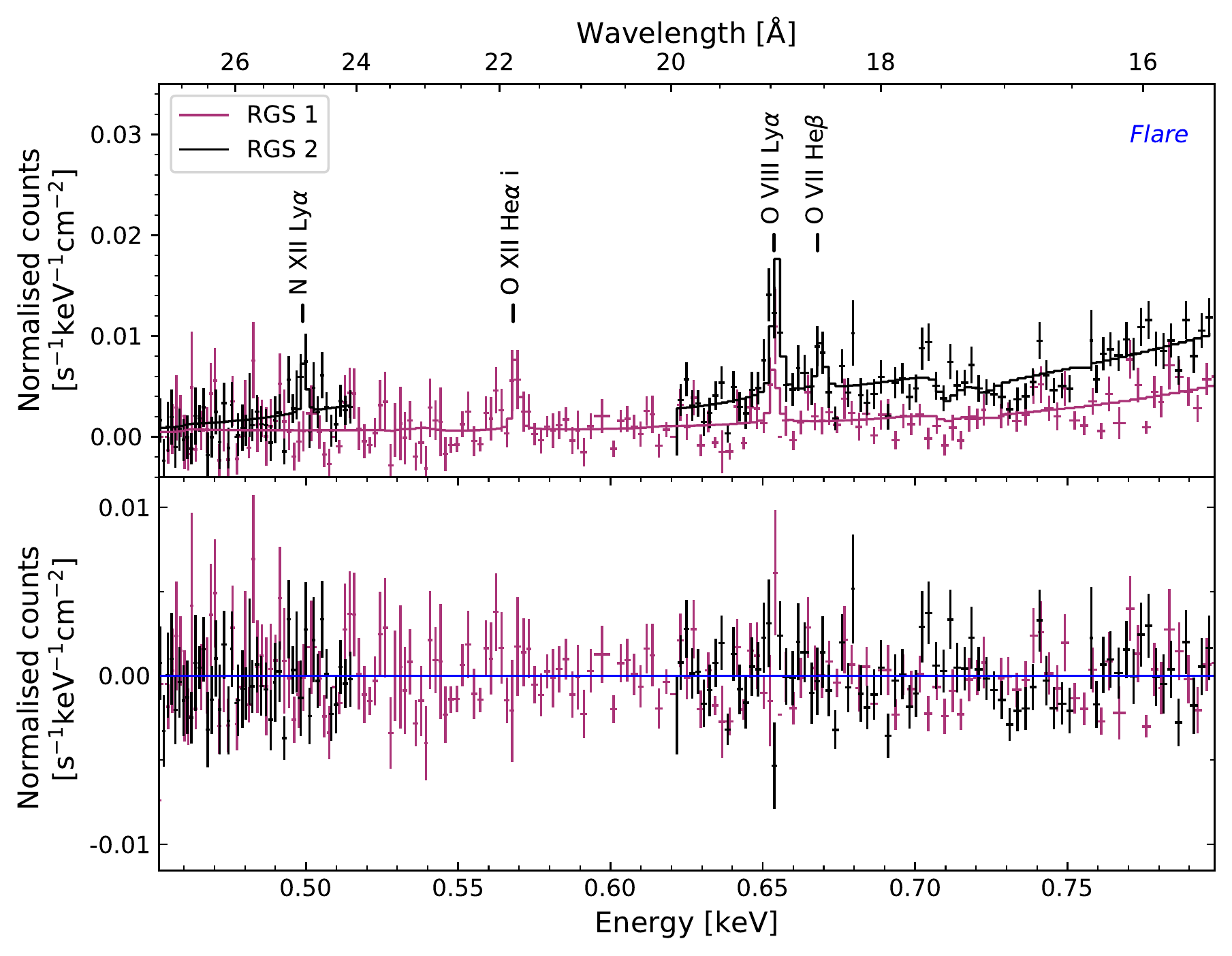}
  \end{subfigure} 
  \begin{subfigure}[b]{0.5\linewidth}
    \includegraphics[width=\textwidth]{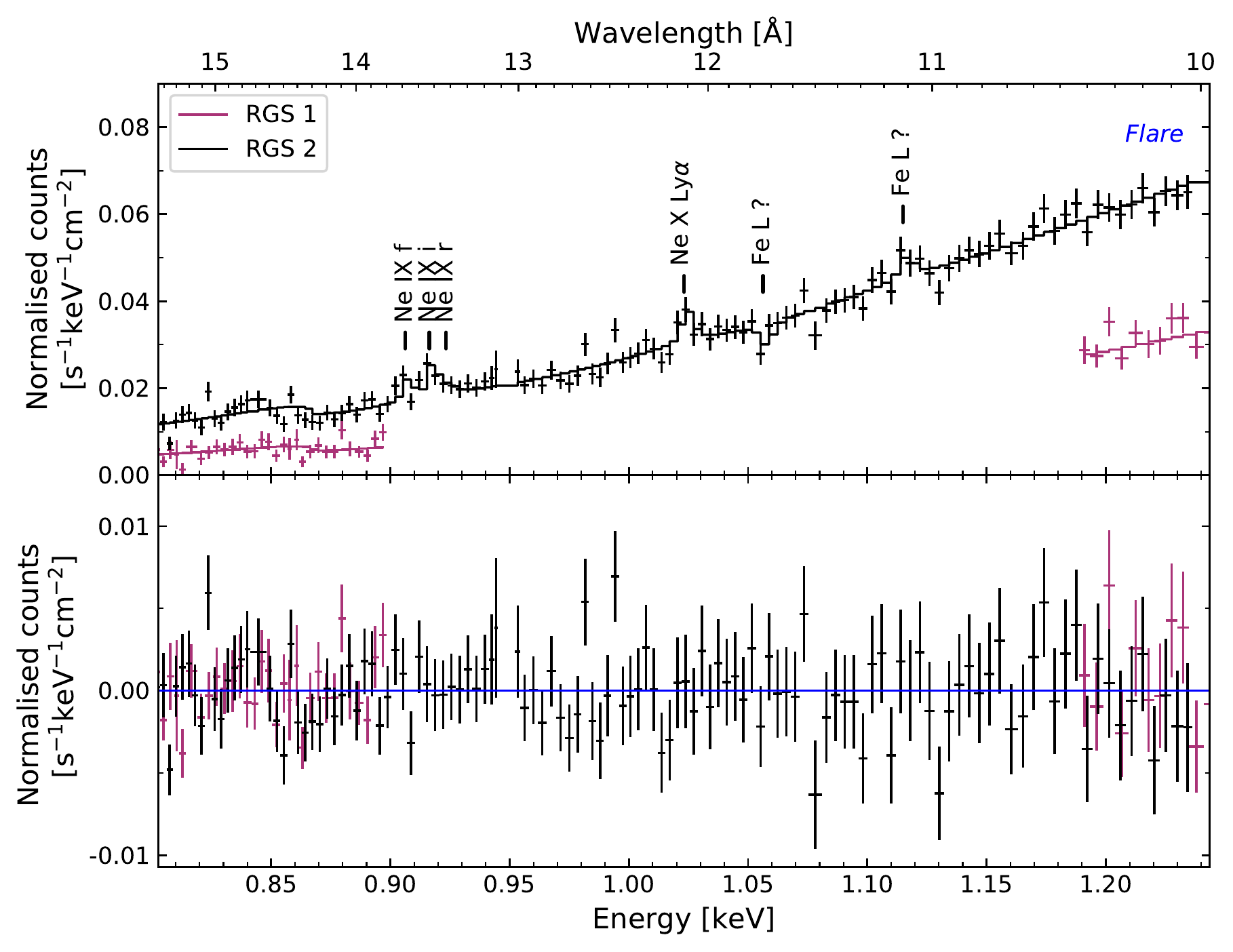}
  \end{subfigure} 
  \begin{subfigure}[b]{0.5\linewidth}
    \includegraphics[width=\textwidth]{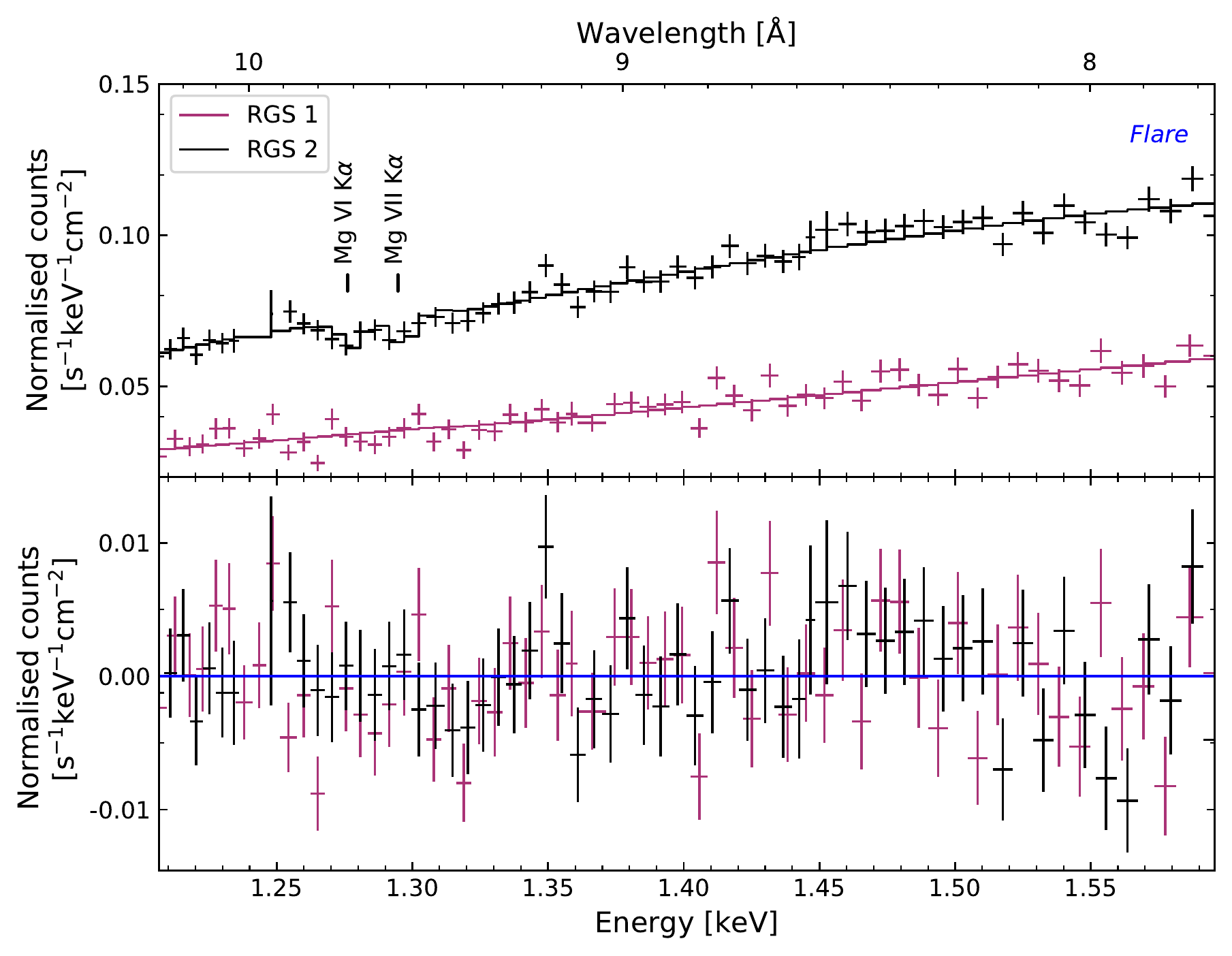}
  \end{subfigure}
  \hfill
  \begin{subfigure}[b]{0.5\linewidth}
     \includegraphics[width=\textwidth]{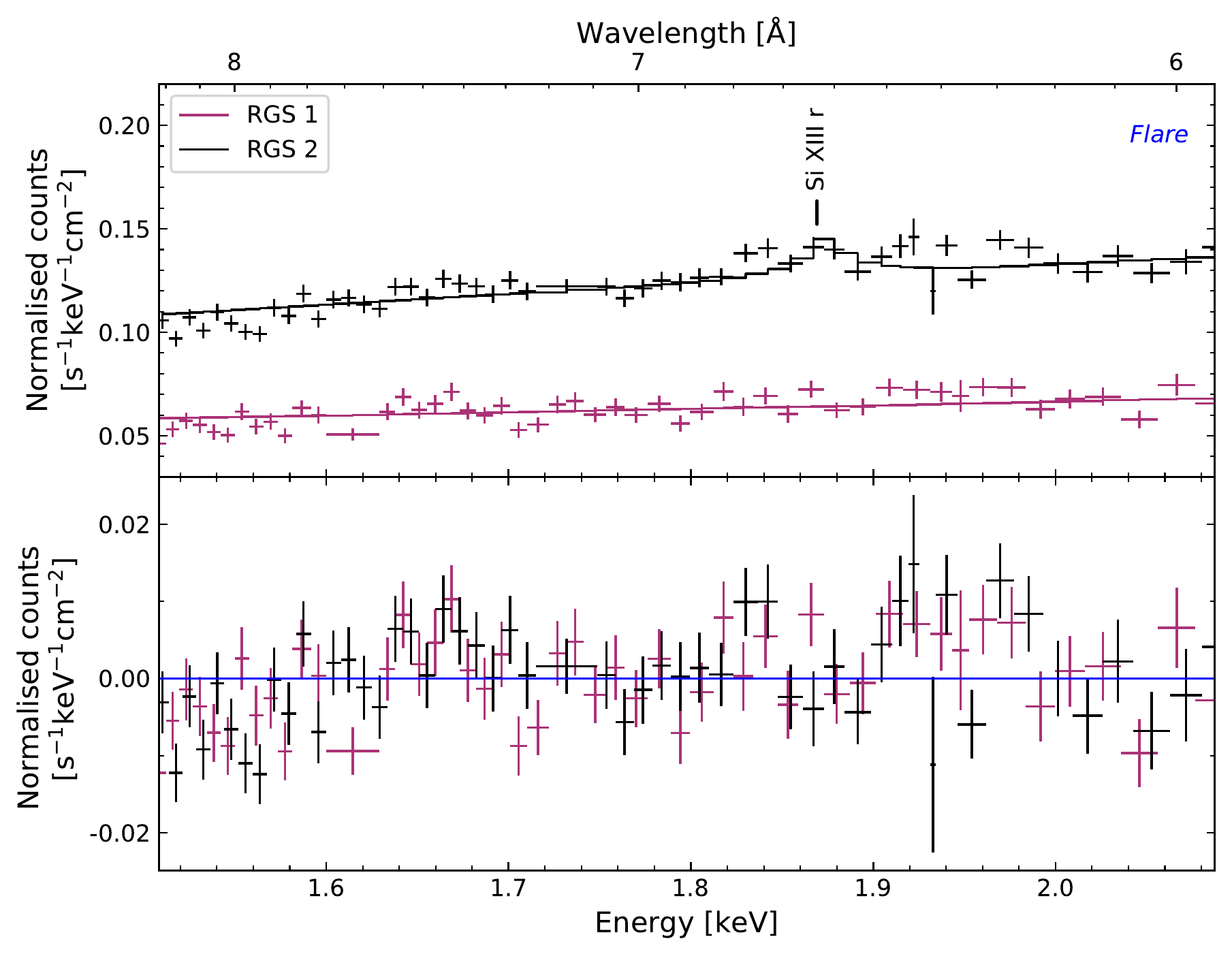}
  \end{subfigure} 
\caption{{\it Upper panel}: RGS~1 and RGS~2 spectrum of the flare phase in four energy bands ({\it crosses}), and best-fit model containing Gaussian components ({\it solid line}). {\it Lower panel}: Residuals in units of data-model. The flux difference between RGS~1 and RGS~2 are attributed to the different coverage of the flare, as shown in Fig. \ref{fig:rgs_lc}.} \label{fig:flare_spectra}
\end{figure*}

\begin{figure*}[htbp]
  \begin{subfigure}[b]{0.5\linewidth}\label{fig:spectra_postflare}
   \includegraphics[width=1\textwidth]{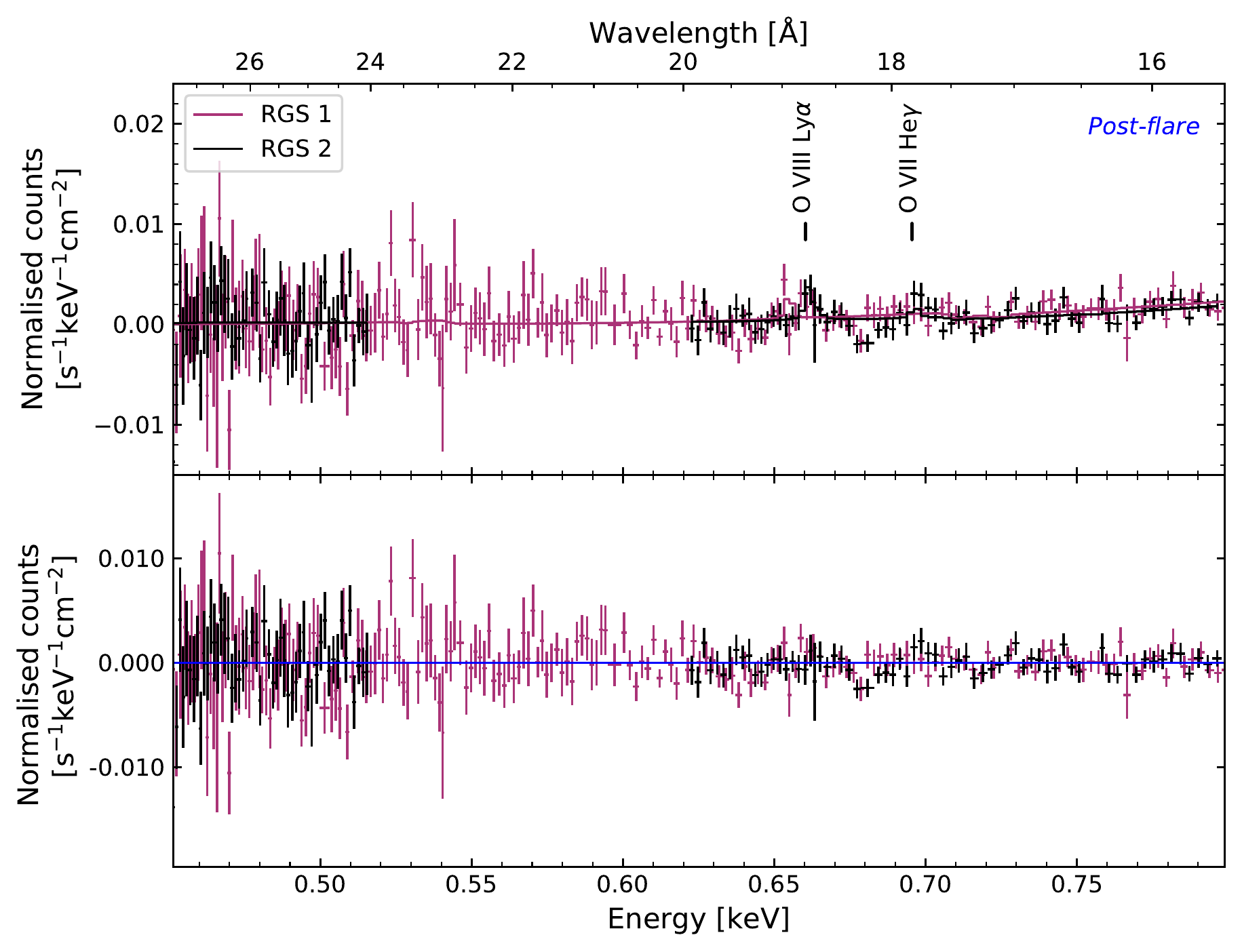}
  \end{subfigure} 
  \begin{subfigure}[b]{0.5\linewidth}
    \includegraphics[width=\textwidth]{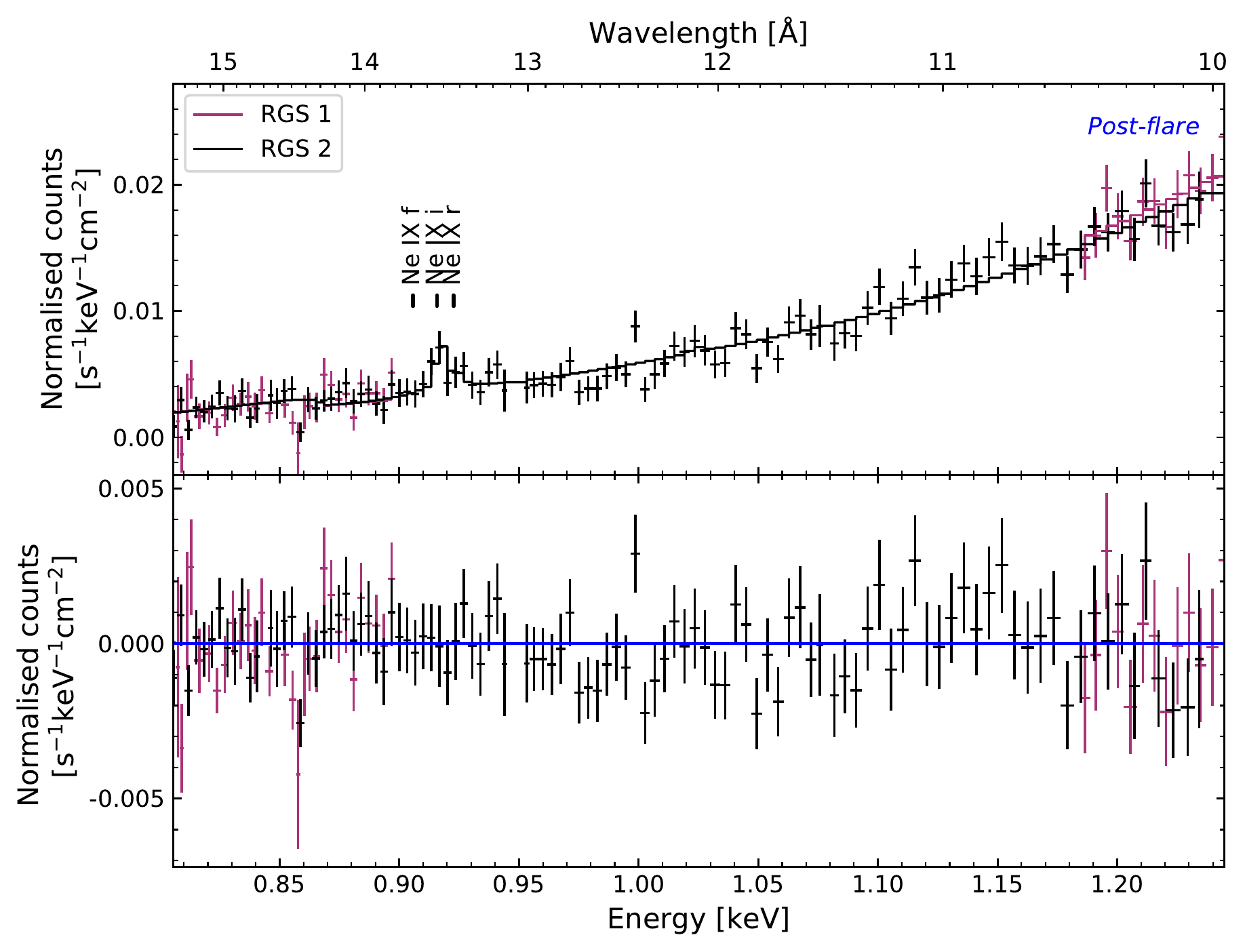}
  \end{subfigure} 
  \begin{subfigure}[b]{0.5\linewidth}
    \includegraphics[width=\textwidth]{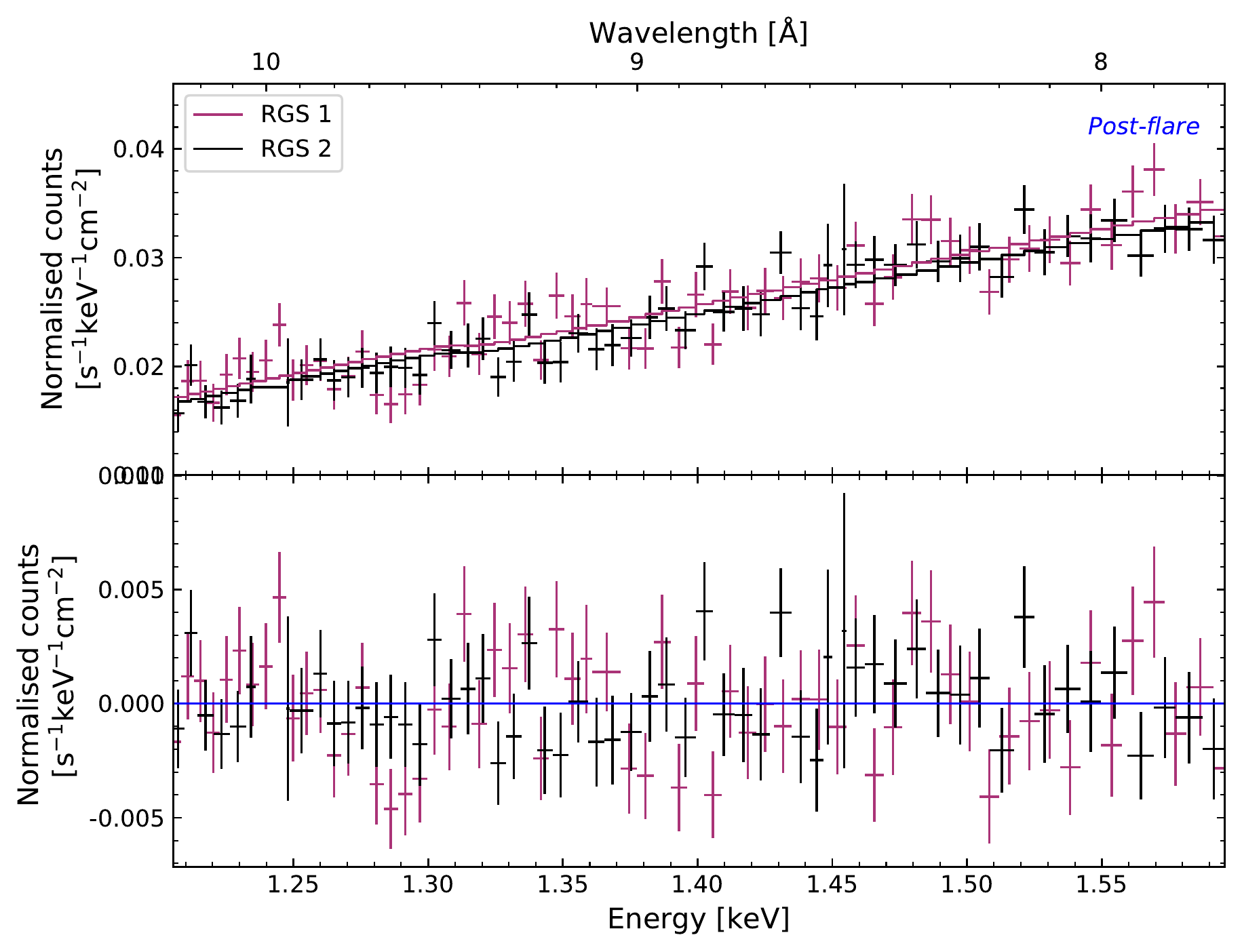}
  \end{subfigure}
  \hfill
  \begin{subfigure}[b]{0.5\linewidth}
     \includegraphics[width=\textwidth]{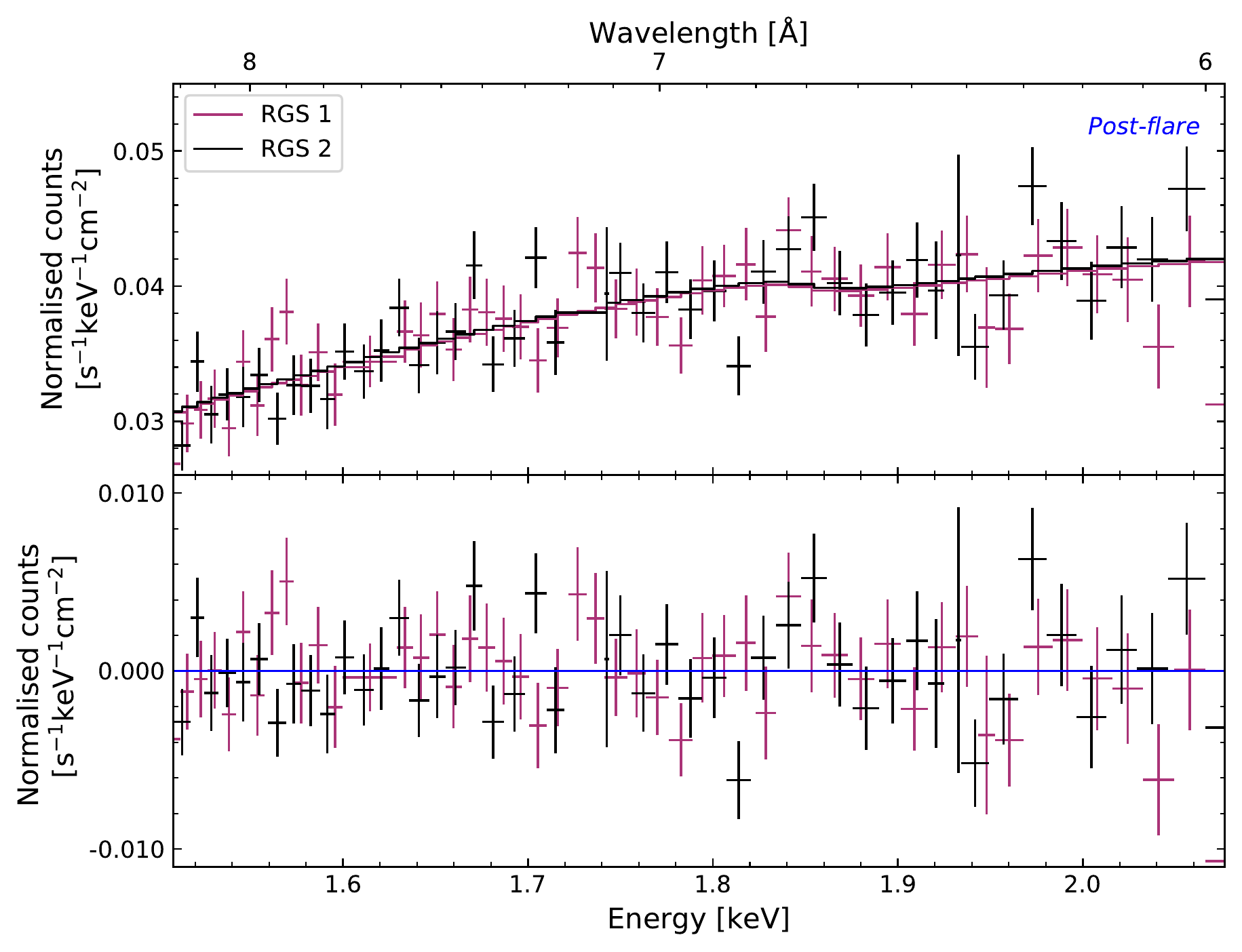}
  \end{subfigure} 
\caption{{\it Upper panel}: RGS~1 and RGS~2 spectrum of the post-flare phase in four energy bands ({\it crosses}), and best-fit model containing Gaussian components ({\it solid line}). {\it Lower panel}: Residuals in units of data-model.} \label{fig:postflare_spectra}
\end{figure*}

Spectra and best fitting models in regions around the detected lines are shown in Fig.~\ref{fig:preflare_spectra} to Fig.~\ref{fig:postflare_spectra}.  
We performed detailed spectral fits only in the regions where our blind line search did yield significant line detections during any of the phases. In particular,  this means that we modeled the 300--650\,eV, 600--800\,eV, 800--1600\,eV, and 1600--2100\,eV ranges in all three phases (pre-flare, flare and post-flare) for RGS~1 and 300--550\,eV, 600--800\,eV, 800--1600\,eV, and 1600--2100\,eV ranges in all three phases for RGS~2 (cf. Tab. \ref{tab:bls_interv}). 

Ideally, each of the spectral lines and RRC in our model should have been absorbed by a separate \texttt{tbabs} component to avoid any assumptions about what part of absorption affects these line-like features; however, as shown in Eq. \ref{eq:model}, they were assumed to be unabsorbed. Due to the relatively narrow energy ranges above and the independence of the line-like features from each other, except for the \ion{Ne}{ix} and \ion{Mg}{xi} triplet complexes as discussed below, we would not have been able to constrain the \texttt{tbabs} components, thus, no ISM absorption was implemented (cf. the \texttt{CLOUDY} model in Eq. \ref{eq:model_cloudy} in  Sect. \ref{sect:cloudy_analysis}).

In summary, we have detected the lines identified as the following transitions: \ion{N}{vii} Ly$\alpha$, \ion{O}{vii} He-$\alpha$, \ion{O}{viii} Ly$\alpha$,  \ion{O}{vii} He-$\beta$, \ion{O}{vii} He-$\gamma$, \ion{Ne}{ix} He-$\alpha$ (f, i, r), \ion{Ne}{x} Ly$\alpha$, \ion{Ne}{x} Ly$\beta$, \ion{Mg}{vi} K$\alpha$, \ion{Mg}{vii} K$\alpha$, \ion{Mg}{xi} He-$\alpha$ (f, r), \ion{Mg}{xii} Ly$\alpha$, and \ion{Si}{xiii} He-$\alpha$ (r) as well as \ion{O}{viii} RRC. The best-fit parameters for all the Gaussian profiles in Eq.\ref{eq:model} detected at a confidence level higher than 90\% (two interesting parameters) are shown in Tab.~\ref{tab:lines} for the time-averaged spectrum  and Tab. \ref{tab:all_lines} for the time-resolved spectra, together with the most likely identification. The detected RRC are listed in Tab.~\ref{tab:rrc} and discussed in Sect.~\ref{sec:rrc}. Finally,  the three transitions with ambiguous identifications are presented in Tab.~\ref{tab:no_id_lines} and discussed in Sect.~\ref{sec:FeL_lines}.

In the case of the observed He-like triplets (\ion{Ne}{ix} and \ion{Mg}{xi}), we fitted them as a combination of three Gaussian components, for which relative intensity was expressed as a function of the $R$ and $G$ parameters, as described in Sect. \ref{sect:rg}. The line energies of each component in the \ion{Ne}{ix} triplet were tied to the \ion{Ne}{x} Ly$\alpha$ line using transition reference energies as published in the literature (see Tab. \ref{tab:all_lines} for references) under the assumption that these lines belong to the same dynamic region and, thus, are affected by the same Doppler shift. For the \ion{Mg}{xi} triplet, the intercombination and forbidden lines were tied to the resonance line in a similar manner.


\begin{table*}[ht]
\centering
\caption{Detected absorption or emission lines (unresolved) in the time-averaged RGS~1 and RGS~2 spectra. }\label{tab:lines}
\renewcommand{\arraystretch}{1.5}
\begin{tabular}{l c c c c c}
\hline \hline

 Ion &  E$_\textnormal{ref}$ & Line energy   &  Normalization & $\Delta C$ & Instrument  \\
& [eV]& [eV] &  $[10^{-5}$ ph\,cm$^{-2}$\,s$^{-1}]$& &  \\\hline

\ion{O}{vii} He-$\alpha$ (i) & 568.7 \tablefootmark{a} & 569.4$^{+0.5}_{-1.9}$& 2.2$^{+1.3}_{-1.4}$ & 10.4 & RGS~1  \\
\ion{O}{viii} Ly$\alpha$ & 654.12 \tablefootmark{b}& 653.8$^{+0.5}_{-1.1}$ & 2.1$^{+1.2}_{-1.1}$ & 18.0 & RGS~1\\

\ion{O}{viii} Ly$\alpha$ & 654.12 \tablefootmark{b}& 653.4$^{+0.5}_{-0.8}$ & 1.6$^{+0.5}_{-0.4}$ & 38.6 & RGS~2\\
\ion{Ne}{ix} (f) & 905.74 \tablefootmark{b} & 904.8$_{-2.2}^{+1.3}$  & 1.8$_{-0.7}^{+0.9}$ & 28.2 & RGS~2  \\ 
\ion{Ne}{ix} (i) & 915.58 \tablefootmark{b} & 916.2$_{-1.4}^{+0.5}$ & 3.7$_{-0.9}^{+1.0}$ & 96.5 & RGS~2 \\

\ion{Ne}{x} Ly$\alpha$  & 1022.60 \tablefootmark{c} & 1022.9$_{-1.5}^{+2.0}$   &  3.4$_{-1.2}^{+1.2}$ & 42.9 & RGS~2 \\
\ion{Fe L}{xxiv} (?) & 1109.379\tablefootmark{d} & 1114.2$_{-4.0}^{+4.6}$   &  2.6$_{-1.4}^{+1.6}$ & 15.1 & RGS~2 \\
\ion{Mg}{xi} (f/i/r) & 1332.11/1344.33/1353.26 \tablefootmark{b} & 1348.1$^{+8.3}_{-12.3}$ & 3.5$^{+2.9}_{-2.9}$ & 31.3 & RGS~2\\
\hline
\ion{Mg}{vi} K$\alpha$& 1276.81\tablefootmark{e} & 1275.4$^{+7.6}_{-10.7}$ & -3.7$^{+2.7}_{-2.0}$ & 6.3 & RGS~2 \\
\ion{Mg}{vii} K$\alpha$ & 1289.95\tablefootmark{e} & 1293.1$^{+2.9}_{-2.7}$ & -6.1$^{+2.5}_{-2.4}$  & 10.7 & RGS~2 \\ \hline

\end{tabular}

\tablefoot{The statistical significance of the lines is denoted by $\Delta C$, i.e., the difference of the model C-statistic value with and without the line. The error bars were calculated for a 90\% confidence interval and two interesting parameters. The line at 1114.24 eV is presumably identified as a \ion{Fe L}{xxiv} line, hence, the question mark (see Sect. \ref{sec:FeL_lines} for discussion).
}
\tablebib{
\tablefoottext{a}{\cite{grant_80}}
\tablefoottext{b}{\citet{drake_88}} 
\tablefoottext{c}{\citet{erickson_77} } 
\tablefoottext{d}{\cite{brown_02}}
\tablefoottext{e}{From \citet{Behar_2002a}; value for \ion{Mg}{vii}~K$\alpha$ averages
 over the two strong transitions listed there.}

}
\end{table*}

\longtab{
\renewcommand{\arraystretch}{2}
\begin{landscape}
\begin{longtable}{l c|c c c|c c c|c c c} 
\caption{Identified narrow features (emission and absorption lines and RRC) and upper limits in the RGS~1 and RGS~2 spectra. }\label{tab:all_lines}\\
\hline\hline
\multicolumn{1}{c}{}&\multicolumn{1}{c}{}&\multicolumn{9}{|c}{Phase}\\ \cline{1-11}
\multicolumn{1}{c}{}&\multicolumn{1}{c}{}&\multicolumn{3}{|c}{Pre-flare}&\multicolumn{3}{|c|}{Flare}&\multicolumn{3}{c}{Post-flare} \\\cline{1-11}
Ion & $E_\textnormal{ref}$ & $E_\textnormal{obs}$ & $N$  &  $\Delta C$  & $E_\textnormal{obs}$ & $N$  &  $\Delta C$ &$E_\textnormal{obs}$ & $N$   & $\Delta C$ \\

 & [eV]& [eV]&  [ph cm$^{-2}$ s$^{-1}]$ &   & [eV]&  [ph \, cm$^{-2}$ s$^{-1}]$  &    & [eV]&  [ph cm$^{-2}$ s$^{-1}]$ &  \\
  &  & &  $\times 10^{-5}$  &    &  &  $\times 10^{-5}$ &    & &  $\times 10^{-5}$ &   \\\hline
\endfirsthead
\caption{continued.}\\
\hline
\multicolumn{1}{c}{}&\multicolumn{1}{c}{}&\multicolumn{9}{|c}{Phase}\\ \cline{1-11}
\multicolumn{1}{c}{}&\multicolumn{1}{c}{}&\multicolumn{3}{|c}{Pre-flare}&\multicolumn{3}{|c|}{Flare}&\multicolumn{3}{c}{Post-flare} \\\cline{1-11}
Ion & $E_\textnormal{ref}$ & $E_\textnormal{obs}$ & $N$  &  $\Delta C$  & $E_\textnormal{obs}$ & $N$  &  $\Delta C$ &$E_\textnormal{obs}$ & $N$   & $\Delta C$ \\

 & [eV]& [eV]&  [ph cm$^{-2}$ s$^{-1}]$ &   & [eV]&  [ph \, cm$^{-2}$ s$^{-1}]$  &    & [eV]&  [ph cm$^{-2}$ s$^{-1}]$ &  \\
  &  & &  $\times 10^{-5}$  &    &  &  $\times 10^{-5}$ &    & &  $\times 10^{-5}$ &   \\\hline
\hline
\endhead
\hline
\endfoot

\ion{N}{vii} Ly$\alpha$ & 500.71\tablefootmark{a} & 499.3 & $< 0.3$ & $-$& 499.3$^{+1.1}_{-1.9}$  & 1.5$^{+1.6}_{-1.2}$  & 7.3 &  499.3 & $< 1.7$ &$-$\\
\hline
\ion{O}{vii} He-$\alpha$ (i)  & 569.10\tablefootmark{b}  & $-$ & $-$ & $-$ &  568.5$^{+2.4}_{-1.2}$  &  3.0$^{+2.7}_{-2.1}$ &  9.8 & 568.52  & $<1.4$ & $-$  \\ 
\hline
\ion{O}{viii} Ly$\alpha_{RGS 1}$  & 654.12\tablefootmark{a} & $-$ & $-$ & $-$ &  653.6$^{+1.1}_{-1.1}$ &  3.0$^{+2.7}_{-2.3}$  & 7.6 & 653.6$^{+6.5}_{-2.1}$  &  1.7$^{+1.4}_{-1.2}$ & 6.5 \\ 
\hline
\ion{O}{viii} Ly$\alpha_{RGS 2}$ & 654.12\tablefootmark{a} &  654.15$^{+0.8}_{-1.3}$& 1.0$^{+0.3}_{-0.3}$ &  18.8 & 653.1$^{+0.9}_{-0.8}$ & 5.9$^{+2.3}_{-2.1}$ &   58.4 & 661.1$^{+2.4}_{-1.7}$ & 1.7$^{+1.5}_{-1.2}$ &  6.0  \\
\hline
\ion{O}{vii} He-$\beta$ & 666.08 $\pm$ 0.01 \tablefootmark{c} & 668.48  & $< 0.7$ & $-$  & 668.6$^{+4.7}_{-1.7}$ & 2.3$^{+1.9}_{-1.6}$ &  6.3 &  E$_\textnormal{\ion{O}{vii}\, He\,$\gamma$}$ - 32.20  & $< 1.7 $ &$-$\\
\hline

\ion{O}{vii} He-$\gamma$  & 698.32 $\pm$ 0.03\tablefootmark{c} & 698.53 & < 0.7  &$-$  &  E$_\textnormal{\ion{O}{vii}\, He\,$\beta$}$ + 32.20  & < 2.4 &$-$ & 697.6$^{+2.8}_{-1.9}$  & 1.5$^{+1.5}_{-1.3}$ & 6.6 \\ 

\ion{Ne}{ix} (f) & 905.74\tablefootmark{d} & $E_{\ion{Ne}{x}\,\textnormal{Ly}\alpha}$ - 116.89 & 1.3$^{+0.3}_{-0.3}$ &  137.4  &  $E_{\ion{Ne}{x}\,\textnormal{Ly}\alpha}$ - 116.89 & 5.2$^{+2.9}_{-3.1}$ &  39.6 &  $E_{\ion{Ne}{x}\,\textnormal{Ly}\alpha}$ - 116.89 & 0.2$^{+0.1}_{-0.9}$ & 18.4\\
\hline
\ion{Ne}{ix} (i) & 915.58\tablefootmark{d} & $E_{\ion{Ne}{x}\,\textnormal{Ly}\alpha}$ - 106.86  & $-$ &  $-$ &  $E_{\ion{Ne}{x}\,\textnormal{Ly}\alpha}$ - 106.86 & $-$  & $-$ & $E_{\ion{Ne}{x}\,\textnormal{Ly}\alpha}$ - 106.86 & $-$ & $-$ \\
 \hline
 \ion{Ne}{ix} (r) & 922.69\tablefootmark{d} & $E_{\ion{Ne}{x}\,\textnormal{Ly}\alpha}$ - 99.85 & $-$ &  $-$ & $E_{\ion{Ne}{x}\,\textnormal{Ly}\alpha}$ - 99.85& $-$ & $-$ & $E_{\ion{Ne}{x}\,\textnormal{Ly}\alpha}$ - 99.85 & $-$& $-$ \\
 \hline

\ion{Ne}{x} Ly$\alpha$ &  1022.60\tablefootmark{a} &1022.0$^{+0.9}_{-0.5}$ &  2.5$^{+0.8}_{-0.7}$ &  68.3 & 1022.4$^{+1.5}_{-1.8}$  & 8.0$^{+6.7}_{-5.6}$ &  11.0 & 1022.5 & < 3.5   &$-$ \\\hline

\ion{Ne}{x} Ly$\beta$ & 1211.85\tablefootmark{a} & 1212.9$^{+4.2}_{-6.2}$ & 1.7$^{+0.9}_{-0.7}$ &  24.6 & 1212.9 & < 10.8  &$-$ & 1212.9 & $<9.8$& $-$ \\
\hline
\ion{Mg}{vi} K$\alpha$ & 1276.81\tablefootmark{e} & 1275.1 & $>-0.3$ &$-$  &  1275.1$^{+4.9}_{-4.8}$  &   -20.0$^{+11.1}_{-9.9}$ & 17.0  & 1275.1& > $-6.5$ & $-$ \\
\hline

\ion{Mg}{vii} K$\alpha$ & 1289.95\tablefootmark{e} & 1293.1 & $>-0.4$ &$-$  & 1293.1$^{+8.1}_{-4.5}$ & $-24.5^{+10.6}_{-11.4}$  &  24.1 & 1293.1 & > $-7.8$ & $-$  \\
\hline

\ion{Mg}{xi} (f)  & 1332.11\tablefootmark{d} & $E_{\ion{Mg}{xi}\,\textnormal{r}}$ - 21.15 & 1.4$^{+0.5}_{-0.9}$ &  104.4  & $E_{\ion{Mg}{xi}\,\textnormal{r}}$ - 21.15  & < 6.6  & $-$ & $E_{\ion{Mg}{xi}\,\textnormal{r}}$ - 21.15  & < 0.3& $-$ \\ 
\hline

\ion{Mg}{xi} (i)  & 1344.33\tablefootmark{d} & $E_{\ion{Mg}{xi} \,\textnormal{r}}$ - 8.93 &  $-$ & $-$   & $-$ & $-$ & $-$ & $-$ & $-$ & $-$ \\
\hline

\ion{Mg}{xi} (r)  & 1353.26\tablefootmark{d} & 1350.2$^{+3.9}_{-3.4}$ &$-$ &  $-$  & $-$   & $-$ & $-$ & $-$ & $-$ & $-$  \\
\hline

\ion{Mg}{xii} Ly$\alpha$  & 1473.46 $\pm \,7.0 \times 10^{-3}$\tablefootmark{a}  & 1475.4$^{+6.7}_{-3.0}$ & 3.6  $^{+1.3}_{-1.1}$ &  52.1  & 1475.4 &  $<35.3$&$-$ & 1475.35 & < 14.6 & $-$ \\
\hline

\ion{Si}{xiii} (r)  & 1864.84 $\pm$ 0.05\tablefootmark{f} & 1856.7$^{+23.5}_{-1.7}$ &   3.4 $^{+1.4}_{-1.4}$ &  6.8 & 1869.1$^{+0.9}_{-0.6}$  & 71.9$^{+17.6}_{-17.6}$ & 15.2  & 1864.84  & $<23.7$ &$-$  \\
\hline

\end{longtable}

\begin{minipage}{\linewidth}
\renewcommand{\footnoterule}{}
\tablefoot{
$E_\textnormal{ref}$ denotes reference energy, $E_\textnormal{obs}$ is detected energy, $N$ is normalization, and $\Delta C$ is the difference in the C-statistic with and without the line in question. The error bars were calculated for a 90\% confidence interval and two interesting parameters.
}
\tablebib{
\tablefoottext{a}{\citet{erickson_77}} 
\tablefoottext{b}{\citet{grant_80}}
\tablefoottext{c}{\cite{engstrom_1995}}
\tablefoottext{d}{\citet{drake_88}} 
\tablefoottext{e}{From \citet{Behar_2002a}; value for \ion{Mg}{vii}~K$\alpha$ averages over the two strong transitions listed there}
\tablefoottext{f}{\citet{hell_16} with an additional systematic uncertainty of 0.13 eV}.

}
\end{minipage}
\end{landscape}    
}

\subsubsection{Radiative recombination continua}
\label{sec:rrc}

At 865.25\,eV, we observed a narrow emission feature in the pre-flare phase spectrum corresponding to the  \ion{O}{vii} RRC, see Tab.~\ref{tab:rrc}.  We modeled this feature with the \texttt{redge} model in \texttt{XSPEC}. 
At 1362.77\,eV, and thus coincident with the \ion{Mg}{xi}-triplet (Tab.~\ref{tab:all_lines}), we expect a further RRC feature, namely the \ion{Ne}{X} RRC. This feature has been previously seen in the eclipse spectra of Vela X-1 using \textsl{Chandra}-HETG \citep{schulz_02}. Similarly, the feature is seen by \textsl{Chandra}-HETG in the highly absorbed Vela X-1 spectra at $\phi_\mathrm{orb}\approx 0.75$ (Amato et al., in prep.).  Although the blind line search does not explicitly find this feature among the Mg triplet lines given the resolution of RGS, we included it in our overall model to assess its possible contribution. We again used the \texttt{redge} model. The  RRC features appeared to be narrow with $kT_\textnormal{e} = 13.9^{+17.6}_{-6.8}$\,eV for \ion{O}{viii} and $kT_\textnormal{e} <12.6$\,eV for \ion{Ne}{x} (Tab.~\ref{tab:rrc}), and both can be significantly detected only during the pre-flare phase.

\begin{table*}[]
\renewcommand{\arraystretch}{1.5}
    \centering
    \caption{Detected RRC features in the RGS~2 spectra. The values in the flare and post-flare phase are unconstraining, and therefore, omitted.  }\label{tab:rrc}
    \begin{tabular}{c c | c c c c}
\hline \hline
&&\multicolumn{4}{c}{Pre-flare} \\\cline{1-6} 
Ion & $E_\textnormal{ref}$ & $E_\textnormal{obs}$ &$kT_\textnormal{e}$ & $N$ &  $\Delta C$  \\
  & [eV]& [eV]& [eV]&  [ph cm$^{-2}$ s$^{-1}]$&   \\ 
  \hline
   \ion{O}{viii} & 871.41\tablefootmark{a} & 868.0 $^{+3.3}_{-7.3}$ & 13.9 $^{+17.6}_{-6.8}$ & 2.4 $^{+1.6}_{-0.9}$ & 45.3   \\
   \hline 
   \ion{Ne}{x} & 1362.20\tablefootmark{a} & 1367.8 $^{+9.3}_{-6.8}$ & < 12.6  & 1.91$^{+1.1}_{-1.0}$ & 18.1\\
   \hline 
\end{tabular}
 \begin{minipage}{\linewidth}
 \renewcommand{\footnoterule}{}
 \tablefoot{$E_\textnormal{ref}$ denotes reference threshold energy, $E_\textnormal{obs}$ is detected threshold energy, $kT_\textnormal{e}$ is plasma temperature, $N$ is normalization, and $\Delta C$ is the the difference in the C-statistic with and without the feature in question. The error bars were calculated for a 90\% confidence interval and two interesting parameters. 
 }
 \tablebib{
 \tablefoottext{a}{\cite{Garcia_65}}
 }
 \end{minipage}    
\end{table*}

\subsubsection{Possible presence of Fe~L transitions}\label{sec:FeL_lines}

In addition to the already discussed features in Tab. \ref{tab:rrc}, the blind line search suggests further weak lines around 1050\,eV, 1067\,eV, and 1115\,eV (Tab. \ref{tab:no_id_lines}).  Previous \textsl{Chandra}-HETG observations of Vela X-1 during eclipse show hints of emission features around these energies, identified with  Fe~L transitions, although the lines were not explicitly modeled \citep{schulz_02}. 

The energies of the lines at 1050\,eV and 1115\,eV are, within the fitted centroid uncertainties, compatible with some Fe L transitions, namely the \ion{Fe}{xxii} B13 line at  11.770\,\AA{} (1053.392\,eV), the \ion{Fe}{xiii} Be2 line at 11.736\,\AA{} (1056.443\,eV), and the \ion{Fe}{xxiv} Li3 line at 11.176\,\AA{} (1109.379\,eV), as measured with EBIT \citep{brown_02}. All three are the strongest observed transitions in the respective ionization stages for a collisionally ionized plasma; we note however, that at least a part of the emission we observe is likely from a photoionized component (Sect.~\ref{sect:pimodels}).
The line at 1067\,eV, on the other hand, is at best compatible with weak unresolved lines in B-like \ion{Fe}{xxii} and F-like \ion{Fe}{xviii} while two slightly stronger lines in Be-like \ion{Fe}{xxiii} (Be3 at 11.702\,\AA{} or 1059.513\,eV and Be4 at 11.458\,\AA{} or 1082.075\,eV) and one line in Li-like \ion{Fe}{xxiv} (Li1 11.432\,\AA{} or 1084.536\,eV) are just outside of the confidence interval. 

For all three features in Tab. \ref{tab:no_id_lines},  the fitted line centroids have comparatively large uncertainties, the improvement of the fit statistics is somewhat marginal, and the fitted intensities are basically consistent with zero. If these lines were associated with L-shell transitions in Fe ions, we would expect to see a larger number of these features due to the richness of L-shell transitions in Fe. 
It is, in principle, possible that these three features are the strongest lines, with others remaining undetected because of the low signal-to-noise ratio. But only the feature at 1050\,eV is constrained in all three flare phases, the other two lines only appear during flare and post-flare, respectively; and the lines at 1050\,eV and 1067\,eV are observed in absorption, while the feature at 1115\,eV is in emission. 
For these reasons, while we cannot rule out the presence of Fe L lines, the data do not support a firm detection either.

\longtab{
\renewcommand{\arraystretch}{2}
\begin{landscape}
\begin{longtable}{l c |c c c|c c c|c c c} 
\caption{Unidentified narrow features, potentially Fe lines, in the RGS~1 and RGS~2 spectra.}\label{tab:no_id_lines}\\
\hline\hline
\multicolumn{1}{c}{}&\multicolumn{1}{c}{}&\multicolumn{9}{c}{Phase}\\ \cline{1-11}
\multicolumn{1}{c}{}&\multicolumn{1}{c}{}&\multicolumn{3}{|c}{Pre-flare}&\multicolumn{3}{|c|}{Flare}&\multicolumn{3}{c}{Post-flare} \\\cline{1-11}
\multicolumn{1}{c}{Ion} & $E_\textnormal{ref}$ & $E_\textnormal{obs}$ & $N$  &  $\Delta C$  & $E_\textnormal{obs}$ & $N$  &  $\Delta C$ &$E_\textnormal{obs}$ & $N$   & $\Delta C$ \\

 & [eV]& [eV]&  [ph cm$^{-2}$ s$^{-1}]$ &   & [eV]&  [ph \, cm$^{-2}$ s$^{-1}]$  &    & [eV]&  [ph cm$^{-2}$ s$^{-1}]$ &  \\
  &  & &  $\times 10^{-5}$  &    &  &  $\times 10^{-5}$ &    & &  $\times 10^{-5}$ &   \\\hline
\endfirsthead
\hline
\endhead
\hline
\endfoot
\hline
\endlastfoot
\multicolumn{11}{c}{RGS~2} \\ \hline
 
\multirow{1}{1.5 cm}{\shortstack[l]{\\ \ion{Fe L}{xxii}\\ \ion{Fe L}{xxiii}}} & \multirow{1}{1.5 cm}{\shortstack[c]{\\ 1053.392\tablefootmark{a} \\ 1056.443\tablefootmark{a}}} & 1047.6$^{+4.0}_{-4.4}$ & $-0.5^{+0.4}_{-0.7}$ &  6.2 & 1056.3$^{+11.5}_{-26.4}$ & $-5.9^{+5.6}_{-5.0}$ & 6.0 &1053.4 &  > $-$4.3   & $-$\\
\hline 
\multirow{1}{1.5 cm}{\shortstack[l]{\\ \ion{Fe L}{xviii}\\  \ion{Fe L}{xxii} \\ \ion{Fe L}{xviii} \\ \\ \\ \\  \\ \ion{Fe L}{xxiv} }}   & \multirow{1}{1.5 cm}{\shortstack[c]{\\ $-$\tablefootmark{b} \\ $-$\tablefootmark{b} \\ 1059.513\tablefootmark{a} \\ 1082.075\tablefootmark{a} \\1084.536 \tablefootmark{a}}} & 1067.6$^{+4.1}_{-6.3}$ & $-0.5^{+0.5}_{-1.0}$& 4.3 & 1067.6  & $> -5.1$ &$-$ & 1067.6  & $>-2.2$  &$-$\\ [30pt]
\hline

\ion{Fe L}{xxiv} & 1109.379\tablefootmark{a} & 1115.8 & < 0.7 &$-$ & 1115.8$^{+6.0}_{-12.0}$ &  6.1$^{+8.7}_{-5.8}$ & 5.0 & 1115.8  & < 6.8  & $-$  \\
\hline\hline

\end{longtable}
\begin{minipage}{\linewidth}
\renewcommand{\footnoterule}{}
\tablefoot{$E_\textnormal{ref}$ denotes reference energy, $E_\textnormal{obs}$ is detected energy, $N$ is the normalization, and $\Delta C$ is the the difference in the C-statistic with and without the line in question. The error bars were calculated for a 90\% confidence interval and two interesting parameters. 
}
\tablebib{
\tablefoottext{a}{\cite{brown_02}}
\tablefoottext{b}{Many possible transition are applicable. See \cite{brown_02} for details.}
}
\end{minipage}
\end{landscape}    
}

\subsection{$R$ and $G$ diagnostic parameters}\label{sect:rg}

The ratios of the normalization of the forbidden ($f$), intercombination ($i$), and resonance ($r$) lines in He-like triplets can be used to probe temperature, density, and ionizing processes in astrophysical plasmas \citep{gabriel_69, porquet_00}. They are defined as $R = f/i$ and $G = (f+i)/r$, where the $R$ parameter depends primarily on the electron density, $n_\textnormal{e}$, while the $G$ parameter is density-independent but shows an electron temperature, $T_\textnormal{e}$, dependence instead. 

To estimate these parameters, we created a user-defined model in \texttt{XSPEC} containing a sum of three Gaussian functions with three parameters: line energy, width, and intensity. The line intensity of the intercombination and resonance lines were expressed as a factor containing $R$ and/or $G$ according to the equations above, while the normalization of the model corresponded to the line intensity of the forbidden line. This approach allows us to directly fit for $R$ and $G$.

\begin{figure*}[htbp]
  \begin{subfigure}[b]{0.5\linewidth}\label{fig:RG_Ne_preflare}
    \includegraphics[width=1\linewidth]{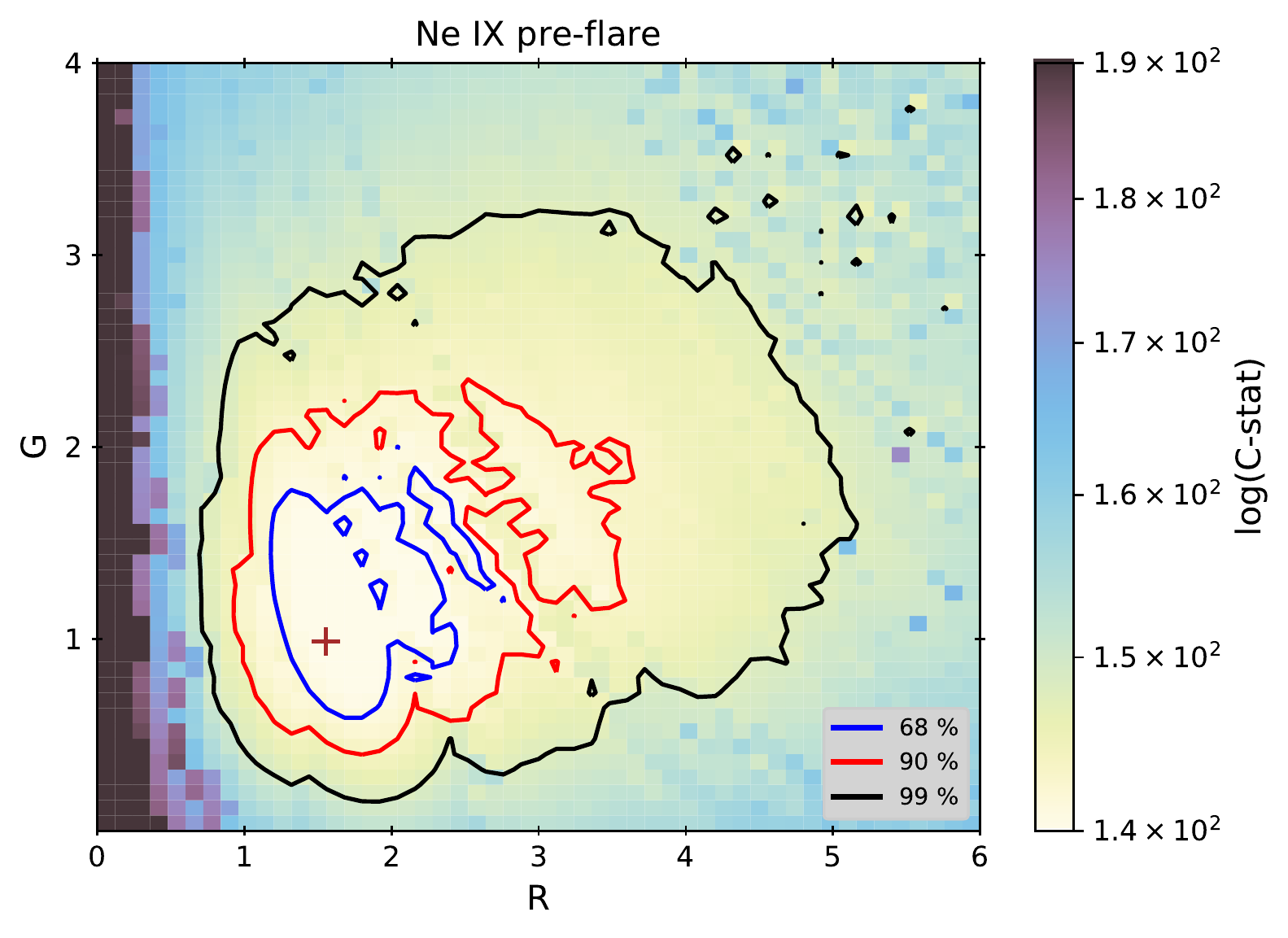} 
  \end{subfigure} 
  \begin{subfigure}[b]{0.5\linewidth}
    \includegraphics[width=1\linewidth]{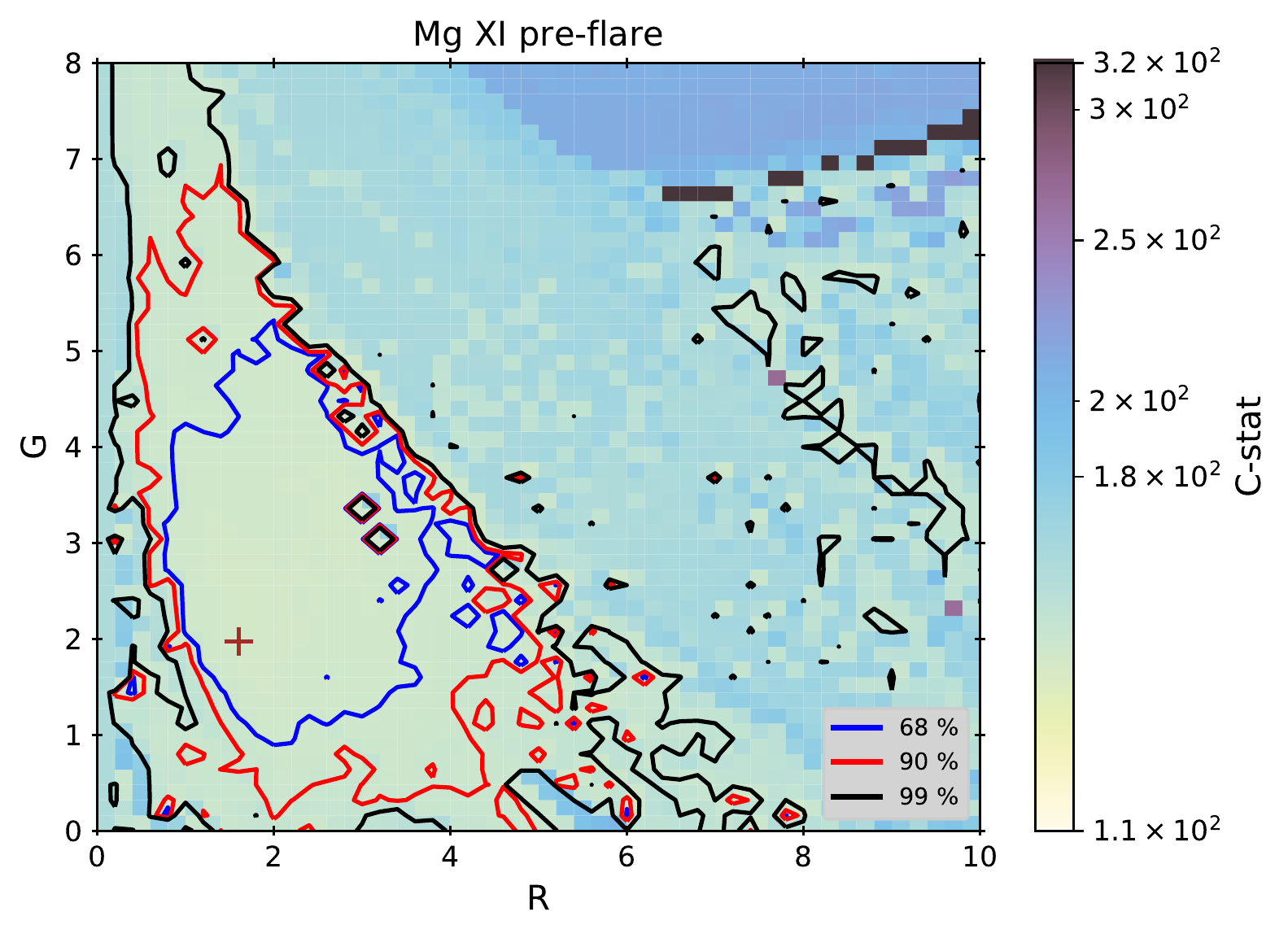}
  \end{subfigure} 
  \begin{subfigure}[b]{0.5\linewidth}
    \includegraphics[width=1\linewidth]{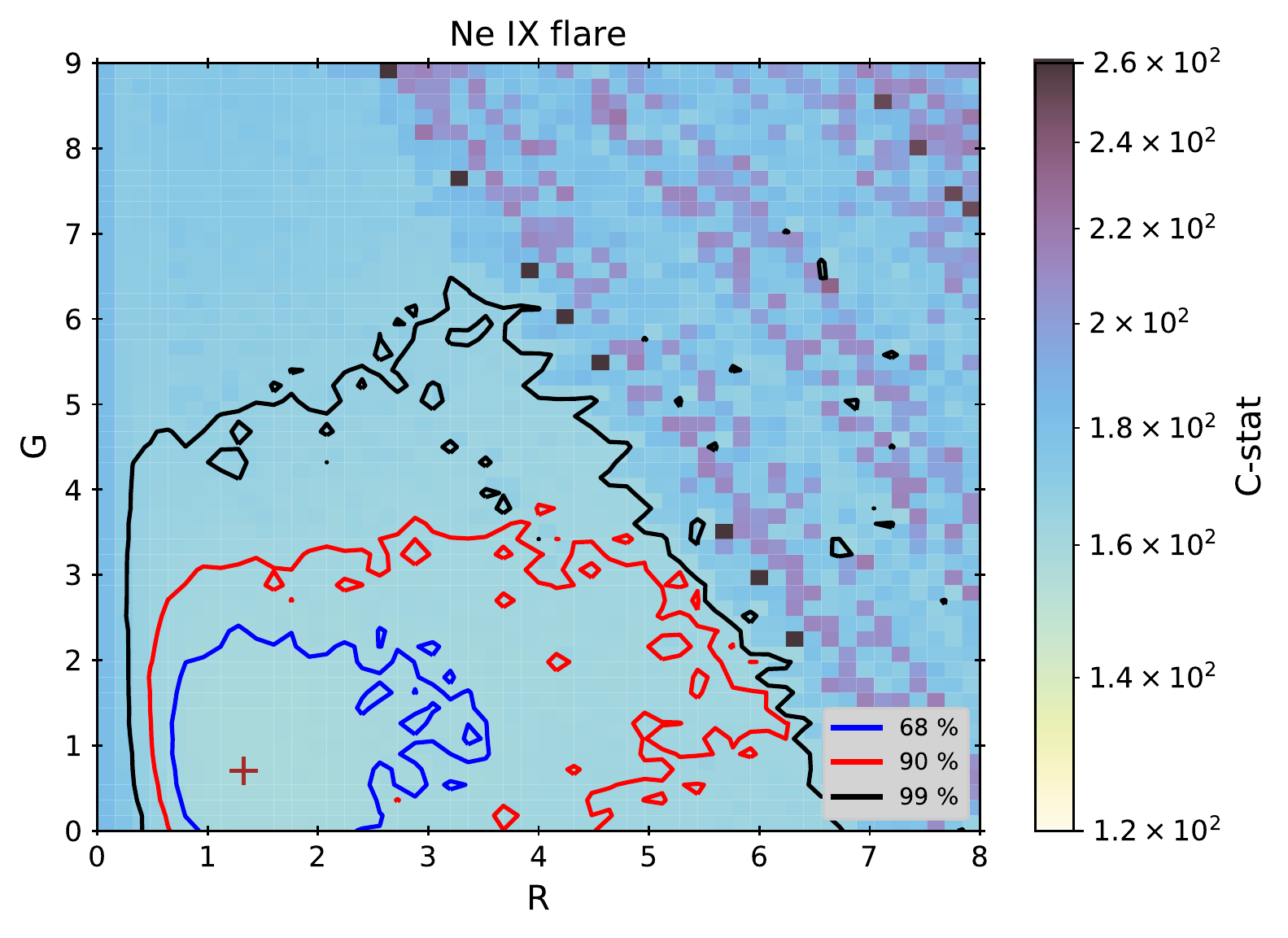}
  \end{subfigure}
  \hfill
  \begin{subfigure}[b]{0.5\linewidth}
    \includegraphics[width=1\linewidth]{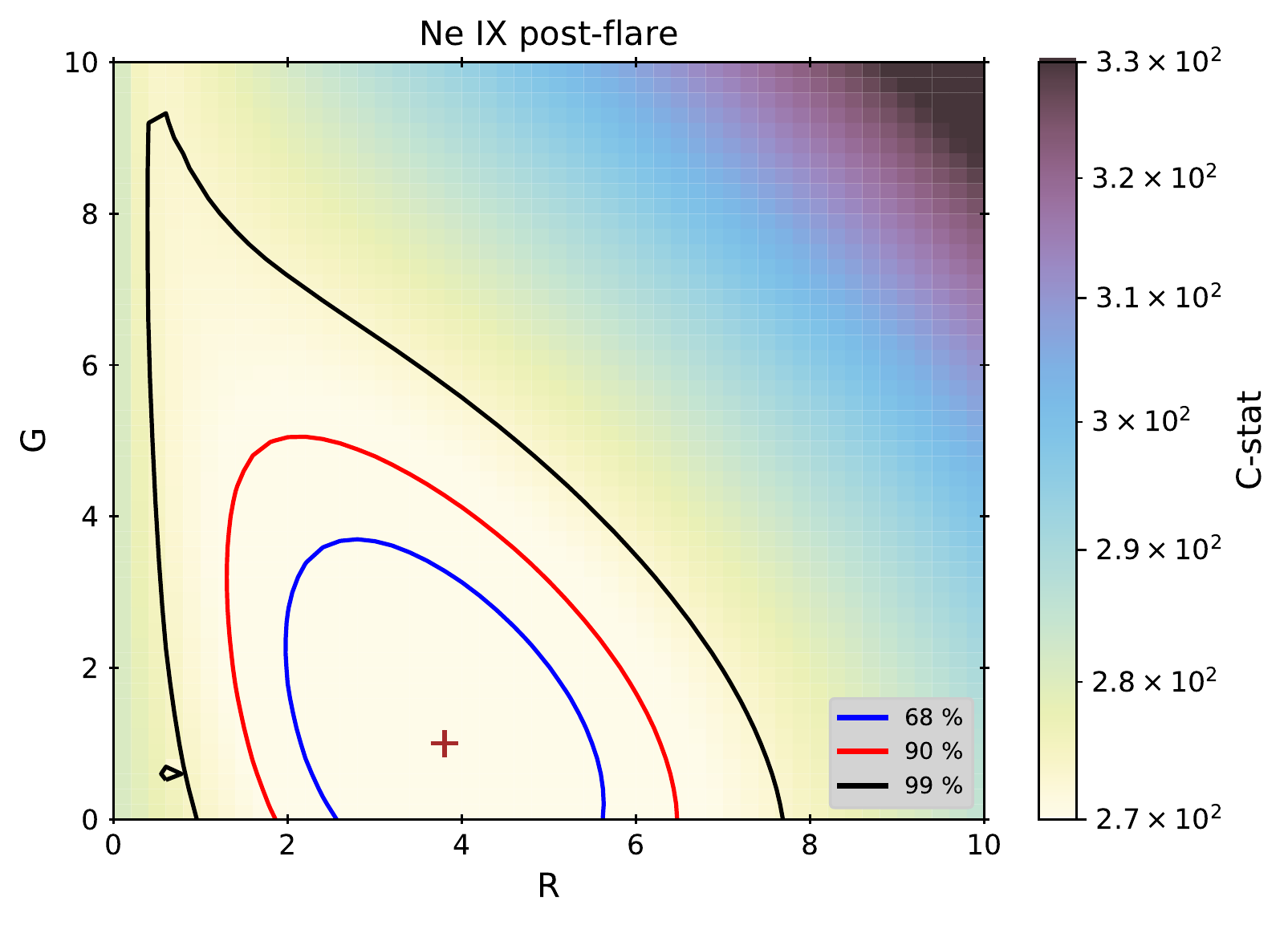}
  \end{subfigure} 
\caption{Iso-C-stat contours in the $R$ vs. $G$ parameter space (50-by-50 grid) for the RGS~2 spectrum. The black, red, and blue contours correspond to 68\%, 90\%, and 99\% confidence intervals, respectively. The crosses denote  best-fit values. \textit{Top left}: \ion{Ne}{ix} triplet in the pre-flare phase. \textit{Top right}: \ion{Mg}{xi} triplet in the pre-flare phase. \textit{Bottom left}: \ion{Ne}{ix} triplet in the flare phase. \textit{Bottom right}: \ion{Ne}{ix} triplet in the post-flare phase. 
} \label{fig:rg_contour}
\end{figure*}

In Fig.~\ref{fig:rg_contour}, we show 2D contour plots of the R and G parameters for the detected triplets in the RGS phase-resolved spectra discussed in this paper. The corresponding values of $n_\textnormal{e}$ and $T_\textnormal{e}$ presented in Table \ref{tab:rg} were taken from \cite{porquet_00} who have calculated the ratios of the diagnostic parameters in fully or partially photoionized plasmas. 

The main results of the analysis can be summarized as follows:

\begin{itemize}
\item 
The $R$ and $G$ diagnostic parameters for the \ion{Ne}{ix}-triplet in the pre-flare phase are constrained at 1 $ \lesssim R \lesssim $ 3.5 and 0.5 $ \lesssim G \lesssim $ 2.5 (here and hereafter we refer to the range within the 90\% confidence interval).

\item
The \ion{Mg}{xi}-triplet was only detected in the pre-flare phase. The contour plot in the top right panel in Fig.~\ref{fig:rg_contour} gives 0.5 $ \lesssim R \lesssim $ 4 , while for the $G$ parameter the range is basically unconstrained ($ G \lesssim 7 $).

\item
For the \ion{Ne}{ix}-triplet in the flare phase, there is a weak correlation between the diagnostic parameters. The confidence interval for the $R$ parameter is rather wide: 0.5 $ \lesssim R \lesssim $ 6, while the $G$ parameter  reaches $ G \lesssim $ 3.5 resembling the pre-flare phase.

\item
In the post-flare phase, we obtained 1.5 $ \lesssim R \lesssim $ 6.5 and $ G \lesssim $ 5 for the \ion{Ne}{ix}-triplet, see the bottom right panel in Fig.~\ref{fig:rg_contour}.  
\end{itemize}

Unsurprisingly, the corresponding ranges of $n_\textnormal{e}$ and $T_\textnormal{e}$ in Tab.~\ref{tab:rg} are rather wide, especially in the flare and post flare phases. Given the pre-flare results, the electron number density varies between $\sim$$10^{10}$ and $10^{12}$ cm$^{-3}$, while the range of the electron temperature is between $\sim$$10^{5}$ and $10^{6}$ K.

In addition, we note that these values should be treated with caution: the results in \cite{porquet_00} do not account for a strong UV radiation field which affects level populations in He-like triplets, i. e., UV pumping. In Sect. \ref{sec:discussion:cloudy}, we show that UV pumping is indeed important in the case of Vela X-1.

\begin{table*}[htbp]
\centering
\caption{Statistical ranges for the R and G parameters given the 90\% confidence interval (red) in Fig. \ref{fig:rg_contour}. 
 }\label{tab:RG}
\renewcommand{\arraystretch}{1.5}
\begin{tabular}{l c c c c c c}
\hline\hline
Phase & Ion & $R$ & $G$  &  $n_\textnormal{e}$ [cm$^{-3}]$ & $T_\textnormal{e}$  [K] & $kT_\textnormal{e}$  [eV]  \\\hline 

Pre-flare & \ion{Ne}{ix} & 1 $ \lesssim R \lesssim $ 3.5 & 0.5 $ \lesssim G \lesssim $ 2.5 & 10$^{11} \lesssim n_e \lesssim 2 \times 10^{12}$  &   10$^6 \lesssim T_e \lesssim  6 \times 10^6$ & 86 $\lesssim kT_e \lesssim$ 517 \\

Pre-flare  & \ion{Mg}{xi} & 0.5 $ \lesssim R \lesssim $4 &  $ G \lesssim 7 $ &  10$^{9} \lesssim n_e \lesssim 5 \times 10^{13}$  &   $2\times 10^5 \lesssim T_e \lesssim  8 \times 10^6$ & 43 $\lesssim kT_e \lesssim$ 689\\

Flare  & \ion{Ne}{ix} & 0.5 $ \lesssim R \lesssim $ 6 &  $ G \lesssim $ 3.5 & 10$^8 \lesssim n_e \lesssim 2 \times 10^{12}$ &  10$^6 \lesssim T_e \lesssim  6 \times 10^6$ & 86 $\lesssim kT_e \lesssim$ 517\\

Post-flare & \ion{Ne}{ix} & 1.5 $ \lesssim R \lesssim$ 6.5&  $ G \lesssim $ 5 & $10^8  \lesssim n_e \lesssim 10^{12}$ & $ 10^5\lesssim  T_e \lesssim  6 \times 10^6$ & 9 $\lesssim kT_e \lesssim $ 517 \\
\hline 
\end{tabular}
\label{tab:rg}
\tablefoot{$n_\textnormal{e}$ denotes electron number density and $T_\textnormal{e}$ stands for electron temperature. These values were taken from \cite{porquet_00}.} 
\end{table*} 
 
\subsection{Photoionization models}
\label{sect:pimodels}

The features detected in the preceding sections, especially in the case of the clear emission lines we observed in the pre-flare spectrum, are indicative of a photoionized medium. We test this assumption through the direct modeling of the pre-flare spectrum with the photoionization model \texttt{CLOUDY} \citep{ferland_2017}.

\subsubsection{Input for photoionization models}
\label{sect:pimodels:input}

We approximated the illuminating continuum, required for the simulation of the photoionized wind medium, as a sum of two components, namely the emission from the neutron star and the emission from HD~77581. A similar approach was employed previously in \citet{Grinberg_2017a}. HD~77581 is dominating the UV continuum and is described by a blackbody, assuming a distance of 2\,kpc,  stellar radius of 31\,$R_\odot$ and a temperature of 20\,000\,K. 

The limited bandpass of the RGS does not allow us to constrain the broadband continuum of the neutron star. However, previous work by \citet{Martinez-Nunez_2014a} characterizes the behavior up to 10\,keV, in particular the X-ray flux of the source in the 0.6--10\,keV range, during our observation. The broadband spectral shape has been best measured with other X-ray telescopes, e.g., with \textsl{NuSTAR} by \citet{Fuerst_2014a}, who find that the observations are best-described with a Fermi-Dirac cut-off \citep[\texttt{FDCUT}; ][]{Tanaka_1986a}:
\begin{equation}
    F(E) = A E^{-\Gamma} \left(1 + \exp
  \left(\frac{E-E_{\mathrm{cut}}}{E_{\mathrm{fold}}}\right) \right)^{-1} \, ,
\end{equation}
with $A$ the overall normalization, $\Gamma$ the photon index,
$E_{\mathrm{cut}}$ the cut-off energy , and $E_{\mathrm{fold}}$ the
folding energy. As in \citet{Grinberg_2017a}, we assumed $\Gamma = 0.99$, $E_{\mathrm{cut}} =20.8$\,keV, and $E_{\mathrm{fold}}
=11.9$\,keV. We then scaled the normalization of the \texttt{FDCUT} such that the flux corresponded to typical fluxes found by \citet{Martinez-Nunez_2014a} in the respective time periods. In particular, for the pre-flare phase a representative value for the unabsorbed 0.6--10\,keV flux is $\sim$3.2$\times10^{-9}$\,erg\,s$^{-1}$\,cm$^{-2}$.

We further estimated the number density of the stellar wind in the vicinity of the neutron star. To do so, we used the mass loss rate, binary separation distance and radius of HD~77581 as described in Sect.~\ref{sec:intro}, while the wind velocity was assumed to be 700\,km\,s$^{-1}$, i.e. the terminal velocity as measured by \citep{gimenez-garcia_16}. Assuming a spherically symmetric propagation of the wind and mass-conservation, we estimated the mass density at the location of the neutron star to $\sim$9$\times10^{-13}$\,kg\,cm$^{-3}$ and the corresponding number density to $\sim$3$\times10^8$\,cm$^{-3}$. To account for possible wind clumping, we then calculated the photoinization models for a range of densities around this value.

\subsubsection{CLOUDY}\label{sect:cloudy_analysis}

We generated a set of photoionized emission models in \texttt{CLOUDY} 17.00 (last described in \citealt{ferland_2017}) for two free parameters: the electron number density, $n_\textnormal{e}$, and the ionization parameter, $\xi$. The electron number density varied between 5.5 $\leq  \log n_\textnormal{e}/\textnormal{cm}^{-3} \leq $ 11.5 in 17 steps, while the ionization parameter varied in 10 steps, such that 0 $\leq \log \xi /\textnormal{erg}\cdot \textnormal{cm}\cdot \textnormal{s}^{-1} \leq $ 4. Additionally, the turbulent velocity of 10 km s$^{-1}$ was taken from \cite{sander_18}, and the hydrogen column density was fixed at 10$^{23}$ cm$^{-2}$, adopting the default \texttt{CLOUDY} solar abundances for a slab geometry. 

To fit the pre-flare RGS~2 spectrum by interpolating between the model grids, we adopted a model defined as:

\begin{equation}\label{eq:model_cloudy}
    F(E) =   \texttt{tbabs}_1 \times \texttt{powerlaw}_1 + \texttt{tbabs}_2 \times  \sum _{i = 1} ^n  \texttt{CLOUDY}_i \, ,
\end{equation}
that is, an absorbed power law (defined by normalization and the photon index, $\Gamma$) and $n$ absorbed additive \texttt{CLOUDY} models (defined by the electron number density, $n_\textnormal{e}$, the ionization parameter, $\xi$, and overall normalization).

\begin{figure*}[htb]
\sidecaption
\includegraphics[width=12cm]{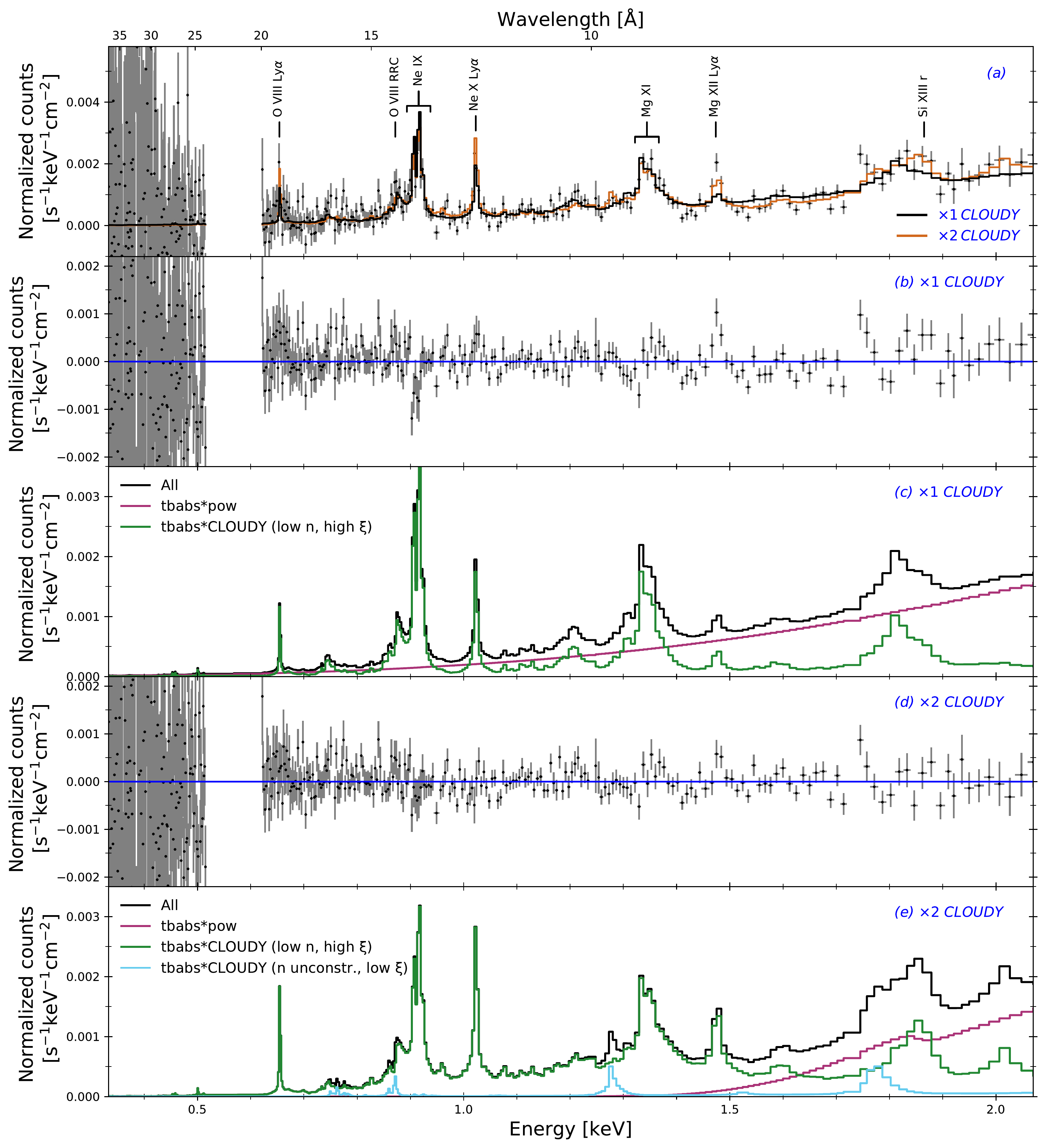}
\caption{RGS~2 pre-flare spectrum fitted with photoionized \texttt{CLOUDY} models. Panel (a): the spectrum and the best fit models with one and two \texttt{CLOUDY} components; panel (b): model residuals  (one \texttt{CLOUDY} component); panel (c): the total model with one \texttt{CLOUDY} component in black and the additive model components in color; panel (d) and (e): similarly as in panel (b) and (c) but for a model with two \texttt{CLOUDY} components. }
\label{fig:cloudy_model_comp}
\end{figure*}

Unlike the model in the phenomenological analysis shown in Eq. \ref{eq:model}, where none of the line-like features were absorbed by \texttt{tbabs}, the $n$ \texttt{CLOUDY}  models are absorbed here. The difference lies in the fit being performed for the entire energy range for \texttt{CLOUDY} and only two \texttt{CLOUDY} parameters ($\log n_\textnormal{e}$ and $\log \xi$) determining the line-like spectral features over the entire energy range; thus, it was possible to constrain the \texttt{tbabs} component.  Panels~(a)~and~(b) in  Fig.~\ref{fig:cloudy_model_comp}, show the fitted model with one \texttt{CLOUDY} component and the corresponding residuals, respectively.

Since several lines were not fitted perfectly, for example, \ion{Ne}{ix} (i), \ion{Mg}{xii} Ly$\alpha$, and Si lines at energies above $\sim$1.7 keV, we added another \texttt{CLOUDY} component with a lower $\log \xi$ value such that both \texttt{CLOUDY} components were modified by the same \texttt{tbabs} model, see Eq. \ref{eq:model_cloudy}. Indeed, this improved the quality of the fit which can be seen from the residuals in panel~(d) in Fig.~\ref{fig:cloudy_model_comp} and from the C-statistic values in Tab.~\ref{tab:cloudy}: $\Delta C/\Delta \nu$ = 62/4. The first component is significantly more dominant as shown in panel~(e) in Fig.~\ref{fig:cloudy_model_comp}, while the contribution of the low ionization component is mostly important at higher energies in the Si region.

\begin{table}[htbp]
\centering 
\renewcommand{\arraystretch}{1.5}
\caption{Best fit values of the photoionized \texttt{CLOUDY} models in the pre-flare phase (RGS~2).}\label{tab:cloudy}
\begin{tabular}{l l r }
\hline\hline
Model & Parameter  & \multicolumn{1}{c}{Value}  \\ \hline 
\multicolumn{3}{c}{$1\times$ \texttt{CLOUDY}}\\\hline
\texttt{tbabs}& $N_\textnormal{H}$ $[10^{22}$ cm$^{-2}]$ &   < 0.2 \\
\texttt{powerlaw} & $\Gamma$ & $-2.9^{+0.5}_{-0.6}$ \\
\texttt{cflux} & $\log F$ [erg cm$^{-2}$s$^{-1}$]& $-11.62_{-0.06}^{+0.05}$ \\ \hline

\texttt{tbabs} & $N_\textnormal{H}$ $[10^{22}$ cm$^{-2}]$ &  1.0$^{+0.2}_{-0.1}$  \\
\texttt{CLOUDY} & $\log n_\textnormal{e}$ [cm$^{-3}]$ & 8.2$^{+0.4}_{-0.2}$  \\
\texttt{CLOUDY} & $\log \xi$ [erg cm s$^{-1}]$ & 3.3$^{+0.6}_{-0.1}$   \\
\texttt{cflux} & $\log F$ [erg cm$^{-2}$s$^{-1}$]& $-$11.0$_{-0.1}^{+0.6}$  \\\hline
C-stat &  & 591\\
$\nu$ &  & 430  \\\hline\hline

\multicolumn{3}{c}{$2\times$ \texttt{CLOUDY}}\\\hline
\texttt{tbabs}& $N_\textnormal{H}$ $[10^{22}$ cm$^{-2}]$ &   12.7$^{+4.8}_{-3.4}$ \\
\texttt{powerlaw} & $\Gamma$ & 5.86$^{+3.95}_{-6.75}$ \\
\texttt{cflux} & $\log F$ [erg cm$^{-2}$s$^{-1}$]& $-11.81_{-0.08}^{+0.07}$  \\ \hline

\texttt{tbabs} & $N_\textnormal{H}$ $[10^{22}$ cm$^{-2}]$ &   0.96$^{+0.07}_{-0.09}$  \\
\texttt{CLOUDY} & $\log n_\textnormal{e}$ [cm$^{-3}]$ & 8.4$^{+0.4}_{-0.4}$  \\
\texttt{CLOUDY} & $\log \xi$ [erg cm s$^{-1}]$ & 3.61$^{+0.03}_{-0.04}$   \\
\texttt{cflux} & $\log F$ [erg cm$^{-2}$s$^{-1}$]& $-11.02_{-0.05}^{+0.05}$ \\\hline

\texttt{tbabs} & $N_\textnormal{H}$ $[10^{22}$ cm$^{-2}]$ &  0.96$^{+0.07}_{-0.09}$   \\
\texttt{CLOUDY} & $\log n_\textnormal{e}$ [cm$^{-3}]$& unconstrained  \\
\texttt{CLOUDY} &  $\log \xi$ [erg cm s$^{-1}]$& 1.7$^{+0.5}_{-0.7}$    \\
\texttt{cflux} & $\log F$ [erg cm$^{-2}$s$^{-1}$] & $-12.3_{-0.4}^{+0.3}$ \\\hline
C-stat &  & 529  \\
$\nu$ &  & 426 \\
\hline \hline
\end{tabular}
\tablefoot{ $N_\textnormal{H}$ stands for hydrogen column density,  $\Gamma$ denotes photon index, $N$ is normalization, $n_\textnormal{e}$ is electron number density, $\log \xi$ denotes ionization parameter, $z$ is red shift, $F$ is flux calculated between 0.33 and 2.1 keV, `C-stat' denotes the C-statistic values, and $\nu$ stands for degrees of freedom. }

\end{table} 

\begin{table}[htbp]
\centering 
\renewcommand{\arraystretch}{1.5}

\caption{Same as \protect{Tab.~\ref{tab:cloudy}} but using photoionized \texttt{XSTAR} models and a collisionally ionized \texttt{XSPEC} model \texttt{vapec}. }\label{tab:xstar}
\begin{tabular}{l l r }
\hline\hline
Model & Parameter  & \multicolumn{1}{c}{Value}      \\ \hline 
\multicolumn{3}{c}{\texttt{XSTAR}}\\\hline
\texttt{tbabs}& $N_\textnormal{H}$ $[10^{22}$ cm$^{-2}]$ &   < 0.2  \\
\texttt{powerlaw} & $\Gamma$ &      $-3.2^{+0.7}_{-0.9}$  \\
\texttt{cflux} & $\log F$ [erg cm$^{-2}$s$^{-1}$]&    $-11.74_{-0.04}^{+0.06}$  \\ \hline

\texttt{tbabs} & $N_\textnormal{H}$ $[10^{22}$ cm$^{-2}]$ &     1.19 $^{+0.05}_{-0.04}$  \\
\texttt{XSTAR} & $n$ [cm$^{-3}]$ & $>2.28\times 10^{11}$  \\
\texttt{XSTAR} & $\log \xi$ [erg cm s$^{-1}]$ &   $2.47^{+0.09}_{-0.07}$   \\
\texttt{cflux} & $\log F$ [erg cm$^{-2}$s$^{-1}$]& $-10.70_{-0.03}^{+0.08}$  \\\hline 
C-stat & & 664  \\
$\nu$ &  &430  \\
\hline \hline
\multicolumn{3}{c}{\texttt{XSTAR + vapec}}\\\hline
\texttt{tbabs}& $N_\textnormal{H}$ $[10^{22}$ cm$^{-2}]$ &   < 3.2  \\
\texttt{powerlaw} & $\Gamma$ &      $-5.2^{+3.5}_{-1.2}$  \\
\texttt{cflux} & $\log F$ [erg cm$^{-2}$s$^{-1}$]&    $-11.73_{-0.06}^{+0.07}$  \\ \hline

\texttt{tbabs} & $N_\textnormal{H}$ $[10^{22}$ cm$^{-2}]$ &     $1.3^{+0.1}_{-0.1}$  \\
\texttt{XSTAR} & $n$ [cm$^{-3}]$ & $>1.6\times 10^{11}$  \\
\texttt{XSTAR} & $\log \xi$ [erg cm s$^{-1}]$ &   $2.41^{+0.19}_{-0.06}$   \\
\texttt{cflux} & $\log F$ [erg cm$^{-2}$s$^{-1}$]& $-10.8_{-0.2}^{+0.2}$  \\\hline

\texttt{tbabs} & $N_\textnormal{H}$ $[10^{22}$ cm$^{-2}]$ & $1.3^{+0.1}_{-0.1}$  \\
\texttt{vapec} & $kT_\textnormal{e}$ [keV]& $0.36^{+0.04}_{-0.05}$   \\
\texttt{vapec} & [Fe/H]& < 0.2           \\
\texttt{cflux} & $\log F$ [erg cm$^{-2}$s$^{-1}$]& $-11.07_{-0.12}^{+0.09}$   \\
\hline 
C-stat & & 559  \\
$\nu$ & & 427 \\
\hline \hline
\end{tabular}
\tablefoot{ $N_\textnormal{H}$ stands for hydrogen column density,  $\Gamma$ denotes photon index, $N$ is normalization, $n$ is gas number density, $\log \xi$ denotes ionization parameter, `C-stat' denotes the C-statistic values, and $\nu$ stands for degrees of freedom.}
\end{table}

\begin{figure*}[htb]
\sidecaption
\includegraphics[width=12cm]{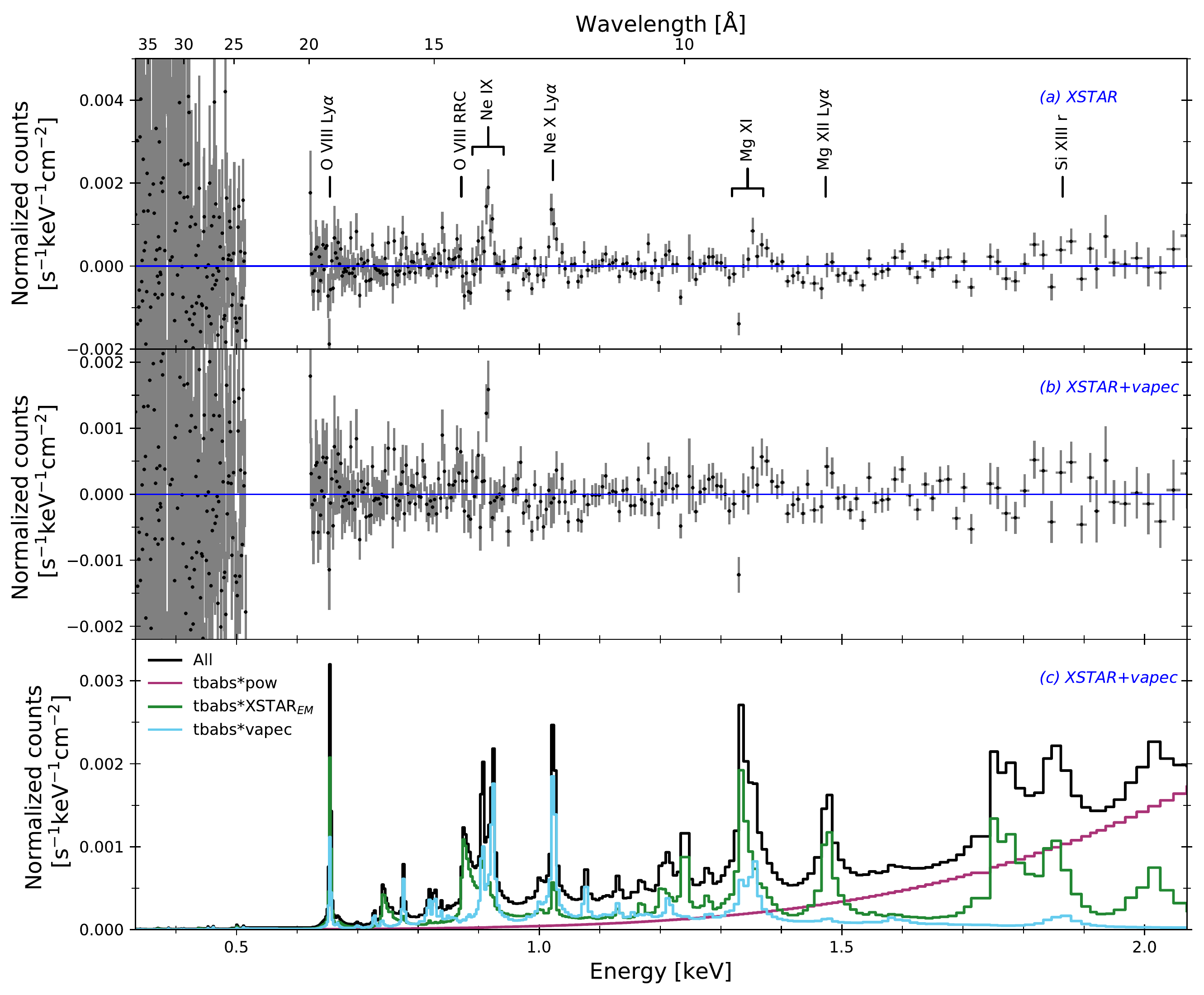}
\caption{RGS~2 pre-flare spectrum fitted with a photoionized \texttt{XSTAR} model and a collisionally ionized \texttt{XSPEC} model \texttt{vapec}. Panel (a): model residuals (one \texttt{XSTAR} component); panel (b): model residuals (\texttt{XSTAR}+\texttt{vapec} components); panel (c): the total \texttt{XSTAR}+\texttt{vapec} model in black and the additive model components in color.}
\label{fig:xstar_model_comp}
\end{figure*}


\section{Discussion}
\label{sect:discussion}

\subsection{Models for wind accretion in HMXBs}

Mass accretion in wind-fed HMXBs occurs through accretion of line-driven stellar winds, in which the compact object is embedded. 
Additionally, there is a possibility of a persistent disk-like structure to be formed around Vela X-1 despite the Roche lobe not being filled by the donor star \citep{elmellah_2019} -- we note, however, that the high magnetic field of the neutron star means that the disk will not extend close to the compact object and an X-ray emitting disk has indeed never been observed in Vela X-1.

The simplest and widely used wind accretion model described by the Bondi–Hoyle–Lyttleton (BHL) theory (see \citealt{edgar_2004} for review) assumes a smooth wind flow to calculate the mass accretion rate onto the compact object. However, line-driven winds of hot stars are known to be subject to instabilities due to velocity perturbations \citep[e.g.,][]{owocki_1988, feldmeier_1997,dessart_2005}. This leads to strong shocks which in turn create dense gas `clumps' (or shells in 1-D models) surrounded by regions of significantly lower densities \citep[see the review by][]{Martinez-Nunez:2017}. Accretion of clumpy winds has been shown to be less efficient than that of homogeneous winds \citep{elmellah_2018}.

The presence of structured winds is well-established in wind-accreting HMXBs. However, the exact physical properties of these structures are under debate as well as the details of the accretion flow \citep{Martinez-Nunez:2017}.  
Some numerical models \citep[e.g.,][]{Blondin:90,Manousakis_2015a} have described the overall flow, including orbital effects and the impact of the X-ray ionization feedback. Others \citep[e.g.,][]{Ducci_2009} have focused on the accretion of individual structures, including also interactions at the magnetosphere and possible changes in the accretion regime \citep{Bozzo_16}. Recently, \citet{elmellah_2018} presented 3D hydrodynamic simulations of the wind in the vicinity of the accretor tracing the inhomogeneous flow over several spatial orders of magnitude, down to the neutron star magnetosphere. However, currently there is no single model treating all the relevant physics across the full range of scales. Thus our measurements cannot yet be tested against a predictive model of the whole accretion process. But they can be used to inform future model developments.

\subsection{Discussion of individual feature detections}

Before a detailed discussion of the individual line feature detection we emphasize that in the time-resolved spectra the detection of emission lines and emission-line-like features such as RRC is much easier in the pre-flare phase, when the continuum is almost fully absent in the RGS range (cf. Fig.~\ref{fig:spectra_all_phases}). The higher continuum level during the flare, but also the post-flare, leads to large uncertainties on line properties and often just upper limits on line contribution.

\subsubsection{Comparison to previous studies of Vela X-1}

We compare the results of our analysis of the time-averaged spectra with the work of \cite{mao_2019}, who have performed an automated line search through the entire RGS archive, including this observation of Vela X-1. They analyzed the observation as a whole and found emission lines at 914.98, 1023.09, 1213.14, and 1483.68\,eV which we also have detected and identified as \ion{Ne}{ix} triplet, \ion{Ne}{x} Ly$\alpha$, \ion{Ne}{x} Ly$\beta$, and \ion{Mg}{xii} Ly$\alpha$, respectively. Additionally, the authors detected emission lines at 1587.92, 1663.94, and 1733.20\,eV, which we have not seen. 
Their automatic approach did not detect the oxygen lines at 569.10 and 654.12\,eV and the \ion{Mg}{xi}, \ion{Mg}{vii}, and \ion{Mg}{vi} lines between 1270 and 1360\,eV that we detect (cf. Tab.~\ref{tab:lines}).  \cite{mao_2019} also observed an emission feature at 824.07\,eV that lacks a clear identification. 

A comparison between \cite{mao_2019} and our results is not straightforward. The two analyses use different statistical criteria to establish the significance of a line, as well as different assumptions on the underlying continuum. Moreover, the \cite{mao_2019} catalog does not provide an assessment of the statistical significance of each line, besides the statistical error on the line normalization. It should be born in mind that the \cite{mao_2019} catalog is the outcome of an automatic, homogeneous analysis on a large number of spectra of completely different classes of objects. Results are not validated on individual observations, because the catalog is intended to be used for statistical studies only. 

As discussed in Sect.~\ref{sect:lines:features}, the line velocities we obtain for the detected lines, both in time-averaged and time-resolved spectra, are consistent with zero and thus not reported. The RGS resolution at energies above $\sim$1\,keV corresponds to line speeds of $\sim$2100--2400\,km/s. For example, line velocities of Si lines in the \textit{Chandra} observations of Vela X-1 in \cite{Grinberg_2017a} have magnitudes of 1450\,km/s or less. Below $\sim$0.4--0.6\,keV, the RGS resolution allows measuring line velocities with magnitudes of $\sim$750--1500\,km/s. However, at lower energies our spectra have a comparatively lower statistical quality.

A plethora of line features has been detected in Vela X-1 previously with different instruments, starting with \textsl{ASCA} \citep{nagase_94}, but none of the other high resolution observations covers a flare as extreme as the one shown here. To our best knowledge, we also present the first firm detection of oxygen and the first detection of nitrogen transitions in the source. In particular, \citet{sako_99} included oxygen into their differential emission measure distribution analysis and reported an upper limit detection of \ion{O}{vii} as well as a possible detection of \ion{O}{viii} RRC.
The \textsl{Chandra}-HETG analyses \citep{schulz_02,Goldstein_2004a,watanabe_06,hell_16,Grinberg_2017a} do not cover such low energies. On the other hand, the HETG observations were able to constrain the weak lines from lower ionization states, especially of silicon, during highly absorbed episodes for which our analysis lacks the resolution.

\subsubsection{Presence of H- and He-like ions}
\label{sec:discussion:HandHE}

The detection of H- and He-like ions is indicative of the presence of hot, highly ionized gas. When modeling the  \ion{Ne}{ix} triplet in all phases, we constrained the centroid energy of the triplet components  relative to that of the Ne{\sc ix} Ly$\alpha$ transition as dictated by the atomic physics, while for the \ion{Mg}{xi} triplet visible in the RGS~2 pre-flare phase, the intercombination and forbidden lines were tied to the resonance line in a similar manner. The supporting assumption implies that these lines have the same Doppler shift and thus belong to the same dynamic system, which seems justified given our data and previous measurements.

 As previously mentioned in Sect. \ref{sect:rg}, the $R = f/i$ and $G = (f+i)/r$ diagnostic parameters obtained from line intensities of He-like atoms can be used as probes for density, temperature, and ionization processes in plasmas. The value of $R$ decreases with increasing electron density, while $G < 4$ in general and $G$ $\sim$1 in particular indicates the presence of a collisionally ionized component \citep{porquet_00,Porquet_2001a}. However, these limits have to be  treated with caution since for accretion-powered pulsars the $G$ parameter can reach low values in a photoionized plasma due to resonance scattering that enhances the resonance line \citep[e.g.,][]{schulz_02, wojdowski_03}. Additionally, the strong UV field of a B star can lead to reduced strength of the forbidden line as the metastable 1s2s$^3$S$_1$ level gets depopulated \citep{gabriel_69,Blumenthal_1972a,Mewe_1978a,Porquet_2001a}. In particular, in Sect.~\ref{sec:discussion:cloudy}
 we test the importance of the star's UV contribution by calculating photoionization models without the stellar contribution to the input continuum. Combined with the large measurement errors on $R$ and $G$ in all cases except perhaps the \ion{Ne}{ix} in the pre-flare phase (Tab.~\ref{tab:rg} and Fig.~\ref{fig:rg_contour}), this precludes us from drawing firm conclusions on the state of the plasma from the He-ratios in this observation. The presence of the lines, however, indicates their potential for future missions with higher throughput and similar resolution at the respective triplet-energies, for example, \textsl{XRISM} and \textsl{Athena} (see Sect.~\ref{sect:future}).

\subsubsection{Low ionization lines}

In the flare phase, we also find absorption lines corresponding to the transitions of fluorescent \ion{Mg}{vi} K$\alpha$ and \ion{Mg}{vii} K$\alpha$ indicating the presence of a less ionized component along the line of sight. \cite{Grinberg_2017a}, who analyzed Vela X-1 at $\phi_\textnormal{orb}$ = 0.25 with \textit{Chandra}-HETG, detected these transitions during the brighter part of their observation as well. Variable absorption has also been previously detected in the high mass X-ray binary Cygnus X-1  \citep{Hirsch_2019a}. An explanation for the presence of such less ionized material would be denser, colder regions in the wind, such as wind clumps that are expected to be present in the wind of O/B giants \citep[e.g.,][]{Oskinova_2012a,Sundqvist_2013a} or parts of the accretion structure \citep[e.g.,][]{Blondin:90,Manousakis_2015a,elmellah_2018}.

\subsubsection{Line response to changes in the continuum}

\begin{figure}[htb]
\centering
\includegraphics[width=0.5\textwidth]{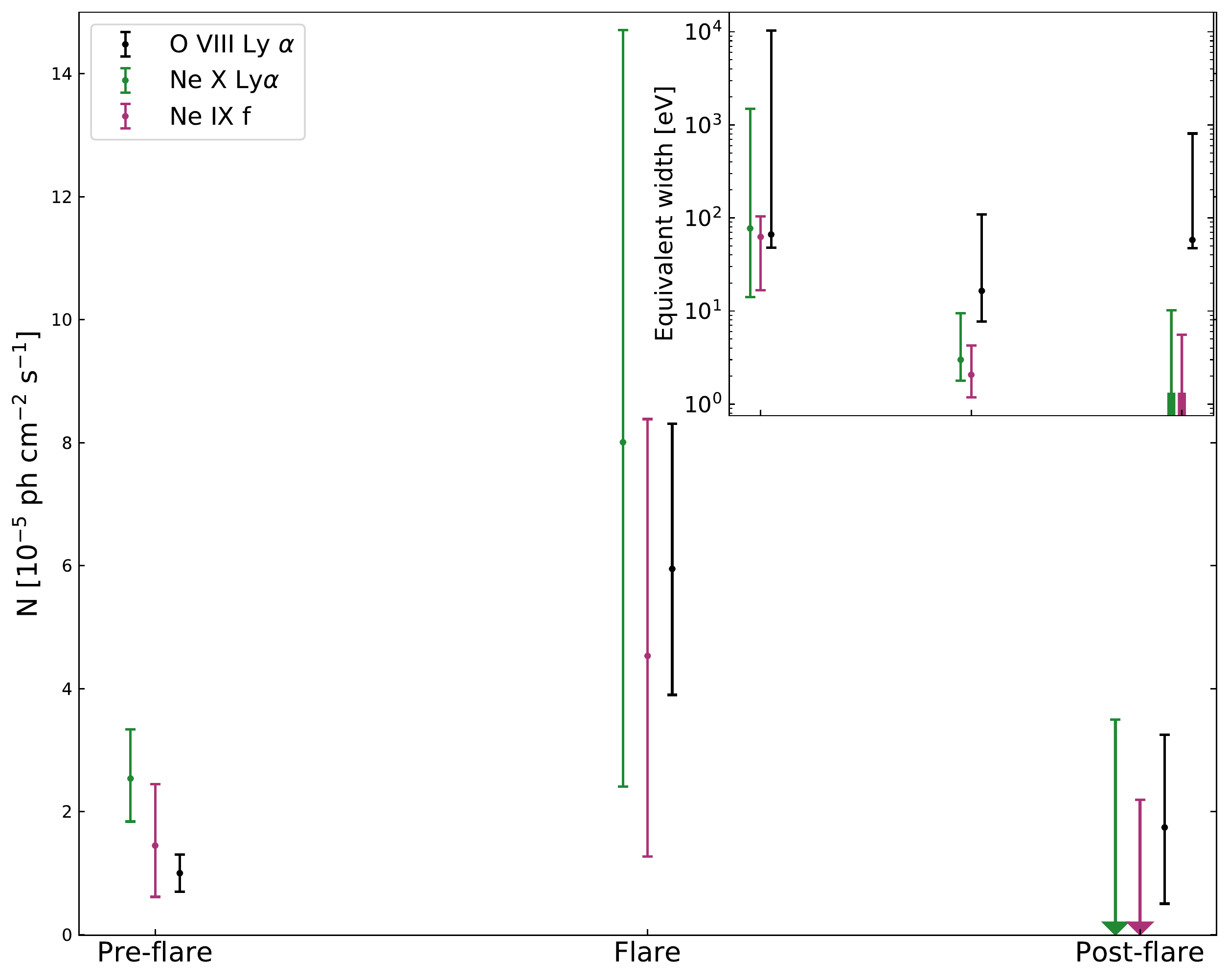}
\caption{Main plot: Time evolution of the line intensity of the \ion{O}{viii} Ly$\alpha$, \ion{Ne}{ix} f, and \ion{Ne}{x} Ly$\alpha$ lines during the pre-, during and post-flare phases with uncertainties (90 \% confidence interval for two interesting parameters). Inset plot: Similarly as above, the time evolution of the equivalent width.} 
\label{fig:ne_o_lines}
\end{figure}

To study potential spectral line changes due to the flare, we investigated the line intensity of \ion{O}{viii} Ly$\alpha$, \ion{Ne}{ix} f, and \ion{Ne}{x} Ly$\alpha$ in the three phases, as shown in Fig. \ref{fig:ne_o_lines}. To do so, we modeled the lines of the \ion{Ne}{ix} triplet as three separate Gaussian components using the build in {\tt XSPEC} function and not as a triplet (cf. Sect. \ref{sect:lines_analysis}). We also investigated the \ion{Ne}{ix} (r) line in the same manner, but we do not include this result in Fig. \ref{fig:ne_o_lines} because the line intensity in this case was consistent with zero in the flare and post-flare phases. All three lines shown follow the same trend reaching an intensity peak during the flare and reducing afterwards.
This trend is consistent with what is seen for the other lines as listed in Tab.~\ref{tab:all_lines}. The concurrent trend of the intensity of the He- and H-like Ne transitions increasing during the flare could be explained by more material becoming ionized as the irradiation of the wind increases.

In the analysis of the \textit{EPIC}-pn data of our observation, \cite{Martinez-Nunez_2014a} saw a peak in Fe K$\alpha$ line flux during the flare, that is, a similar behavior of the line flux as we observe here for other lines. The measured energy of the Fe K$\alpha$ line is compatible with ionization stages of iron up to \ion{Fe}{xviii}, and thus with iron ions that could be produced in the same plasma that creates the lines we observe with RGS. We, in particular, check that this is indeed the case by looking at the iron ionic abundances corresponding to the best fit photoionization models in Sect.~\ref{sect:pimodels}.

In the inset plot in Fig. \ref{fig:ne_o_lines} we also show the evolution of the equivalent width (EW) throughout the flare. We provide these measurements for convenience of comparisons only and in particular emphasize that given the RGS energy range and the strongly varying absorption, the true uncertainties on the continuum and thus on the intrinsic EWs are impossible to estimate with the data we have.

\subsubsection{RRC features}

RRC features arise as free electrons recombine with ions in a plasma. Since the kinetic energy of the electrons is not quantized, but is determined by the Maxwellian speed distribution, its excess w.r.t. the threshold energy of the atomic level will be transferred to the emitted photon. The intensity of an RRC feature is proportional to the electron density, while its width depends on electron temperature. In an overionized plasma, where the electron temperature is much lower than the ionization temperature, the RRC features are expected to be rather narrow and line-like compared to broad RRC in collisional plasmas.

The detected \ion{O}{viii} RRC at 865.25\,eV in the pre-flare phase corresponds to $kT_\textnormal{e} = 13.9^{+17.6}_{-6.8}$\,eV and $T_\textnormal{e} \sim$1.5$\times 10^5$\,K. \cite{schulz_02} analyzed Vela X-1 in eclipse with \textit{Chandra}-HETG and found a similar value,  $kT_\textnormal{e} \sim$10$ \pm 2$ eV and $T_\textnormal{e}\sim$1.2$\times 10^5$ K, using the \ion{Ne}{x} RRC at 1363.52\,eV. In our case, the \ion{Ne}{x} RRC feature cannot be resolved from the \ion{Mg}{xi} because of RGS's resolution, and we obtain a reasonable fit using three Gaussian lines to represent the He-like Mg-triplet only. If we include the RRC into our model, we only obtain an upper limit $kT_\textnormal{e} \leq$12.6\,eV and $T_\textnormal{e} \leq$1.5$\times 10^5$\,K (Tab. ~\ref{tab:rrc}).
The temperatures we obtain from RRC are typical for photoionized plasmas.

We note that, in principle, we should be able to detect a shift in the ionization balance that would allow us to constrain the density independently by measuring the change in the intensity of the \ion{O}{viii} RRC. Given that the timescale of the flare is $\sim$$10^{4}$~sec (see Fig. \ref{fig:rgs_lc}), and assuming that recombination timescales are of the order  $\sim$$10^{12}/(n_\textnormal{e}\cdot\textnormal{cm}^{-3}$)~sec, with the electron density  $n_\textnormal{e} = 10^8$ cm$^{-3}$ (\texttt{CLOUDY} results in Tab \ref{tab:cloudy}), these timescales become comparable. This would be valuable since the density sensitivity of the n = 1--2 He-like ion spectra is likely to be compromised by the UV radiation field of the companion star. Unfortunately, the  quality of the spectra in this work does not allow us to measure such changes.

\subsection{Photoionization modeling}

\subsubsection{\texttt{CLOUDY} modeling}
\label{sec:discussion:cloudy}

As previously mentioned in Sect.~\ref{sect:cloudy_analysis}, one \texttt{CLOUDY} component is sufficient to fit the emission lines in the RGS~2 pre-flare spectrum (see Fig.~\ref{fig:cloudy_model_comp}). Nevertheless, two \texttt{CLOUDY} components manage to fit the data better, which is unsurprising given our expectation of a complex structure of the clumpy stellar winds in Vela X-1 \citep[e.g.,][]{sako_99,schulz_02}. From the best fit values in Tab.~\ref{tab:cloudy}, we can conclude that the two \texttt{CLOUDY} components correspond to two different phases of the plasma: the first one being highly ionized ($\log \xi/\mathrm{erg} \cdot \mathrm{cm} \cdot \mathrm{s}^{-1} \approx 3.6$) and the second with a lower ionization ($\log \xi/\mathrm{erg} \cdot \mathrm{cm} \cdot \mathrm{s}^{-1} \approx 1.7$). The highly ionized component is rather thin, with $\log n_\textnormal{e}/\mathrm{cm}^{-2} \approx 8\,$, but our data do not allow to constrain the density of the second, less prominent \texttt{CLOUDY} component. Less ionized components have, however, been interpreted as colder, denser clumps in other works previously, especially such that employed \textsl{Chandra}-HETG observations and could resolve lower ionization stages of elements such as Si and Mg \citep{schulz_02,Grinberg_2017a}. The number density of the component with the higher ionization corresponds to the value we calculate for the average number density of the wind at the location of the neutron star in Sect.~\ref{sect:pimodels:input}. If there are indeed two components present and the wind is structured with more and less dense regions, a natural identification would be for the thinner component to be more ionized than the denser.

We remind the reader to not take the values of the photo index in the power laws in Tables ~\ref{tab:cloudy} and \ref{tab:xstar} literally. As we mentioned in Sect. \ref{sect:continuum}, our model of the continuum does not constrain its physical properties.

In addition, we also investigate the effects of the UV radiation on the produced models. To do so, we have created an additional set of \texttt{CLOUDY} models for the same input parameters as above with one exception: the illuminating continuum needed for the simulation now did not contain the blackbody contribution from the supergiant (cf. Sect.~\ref{sect:pimodels:input}). The obtained best fits had a significantly poorer quality than the model with the UV contribution ($\Delta$C-statistic $\simeq$150 for the model with two \texttt{CLOUDY} components) and, among others, predicted too strong forbidden lines of the \ion{Ne}{ix} and \ion{Mg}{xi} triplets. This clearly indicates that UV line pumping is important for such simulations, as expected (cf. also Sect.~\ref{sec:discussion:HandHE}).

Ions at significantly different ionization stages found together in the spectrum of Vela X-1 in this work (Mg) and other observations, for example, in the \textit{Chandra} observations in \cite{Grinberg_2017a} (Mg and Si), reveal the complex structure of this system. Describing it as a combination of a small number of discrete photoionization components is likely to be an over-simplification of a rather complex astrophysical system, and more sophisticated plasma models that, for example, can simulate multi-phase plasmas, are needed for these purposes.

\subsubsection{\texttt{XSTAR} modeling}

In addition to the photoionized \texttt{CLOUDY} models, we also address \texttt{XSTAR} modeling. Comparing results from the two codes is crucial to check for the consistency of our results but also to cross-check the codes.

Similarly as for \texttt{CLOUDY}, we allowed two free parameters, the gas density, $n$, and the logarithm of the ionization parameter, $\log \xi$, when generating photoionized models with \texttt{XSTAR}. However, here $n$ varied between 3$\times$$10^{7} \leq n / \mathrm{cm}^{-3} \leq 3$$\times 10^{11}$ in 10 steps. We used the same values for the turbulent velocity from \cite{sander_18}, assumed solar abundances and spherical geometry of the nebula as well as fixing the column density  at 10$^{23}$ cm$^{-2}$.

To obtain an optimal fit implementing \texttt{XSTAR} models, we have tested different combinations of additive emission  and multiplicative absorption \texttt{XSTAR} models as well as combinations of \texttt{XSTAR} with models of optically-thin, collisionally ionized plasma. For the latter component we used {\tt vapec} \citep{smith_2001}, allowing us to fit separately and independently the elemental abundances. 

In panels~(a)~and~(b) in Fig.~\ref{fig:xstar_model_comp} we present the results for the two simplest combinations of models that have produced the best results. Panel~(a) in Fig.~\ref{fig:xstar_model_comp} shows the residuals of the first \texttt{XSTAR} model (pre-flare RGS~2 spectrum) which consists of a power law continuum absorbed by \texttt{tbabs} with a single additive \texttt{XSTAR} emission model absorbed by a separate \texttt{tbabs} component, similarly to the overall \texttt{CLOUDY} model described in Eq. \ref{eq:model_cloudy}.
In panel~(b) in Fig.~\ref{fig:xstar_model_comp} we show the residuals of the model with an absorbed continuum as above and an additive \texttt{XSTAR} model together with one \texttt{vapec} component, both of them being absorbed by the same \texttt{tbabs} (cf.  Eq. \ref{eq:model_cloudy}).

 It is evident that a single \texttt{XSTAR} component is unable to properly fit the \ion{Ne}{ix} triplet and \ion{Ne}{x} Ly$\alpha$ line in the pre-flare RGS~2 spectrum. The differences in the fit quality are also visible from the C-statistic values in Tab. \ref{tab:xstar}, and the strong contribution of the \texttt{vapec} component at lower energies is clearly shown in panel (c) in Fig.  \ref{fig:xstar_model_comp}. The quality of the fits does not significantly improve if an additional {\tt XSTAR} component is added to the model ($\Delta C/\Delta \nu$ = 0.1/7).
 
 The {\tt vapec} component is highly significant ($\Delta C/\Delta \nu$ = 105/3), provided that its iron abundance is low ($Z_\textnormal{Fe}$ < 0.2). A higher Fe abundance would imply much stronger Fe L emission lines than seen in the data. The best-fit {\tt XSTAR} component requires a higher gas density ($n > 2 \times 10^{11}$~cm$^{-3}$) and an intermediate ionization parameter ($\log \xi \simeq 2.4 \, \mathrm{erg} \cdot \mathrm{cm} \cdot \mathrm{s}^{-1}$) with respect to the {\tt CLOUDY} best-fit components (compare Tab. \ref{tab:cloudy} with Tab. \ref{tab:xstar}). One should also note that the estimated temperature of the collisionally-ionized component $kT_\textnormal{e} = 0.36$~keV corresponds to a temperature of the order of 10$^6$~K, which would imply highly ionized Fe ions, for example, \ion{Fe}{xxv}, \ion{Fe}{xxvi}. The Fe K$\alpha$ line from \cite{Martinez-Nunez_2014a}, on the contrary, is consistent with Fe ionized up to \ion{Fe}{xviii}.

The difference in the best-fit parameters between {\tt CLOUDY}- and {\tt XSTAR}-based fit is puzzling. Observationally, it is driven by the fact that {\tt XSTAR} models do not properly fit the Ne {\sc xi} triplet in the Vela~X-1 spectrum, in particular the enhanced ratio between the resonant and the forbidden component, as well as the intensity of the Ne{\sc x} Ly$\alpha$. Similar issues are apparent in the fit of the Mg triplet. Since the main difference between \texttt{CLOUDY} and {\tt XSTAR} in the fit of these spectra seems to be focused on the relative importance of the resonant lines in the He-like triplets, it is likely due to the different assumed geometries and treatment of resonant scattering. We do not have any clear evidence for the collisional phase of the wind plasma, which seems needed because {\tt XSTAR} cannot reproduce the resonant lines at the expenses of the iron abundance, while \texttt{CLOUDY} finds an easier self-consistent solution. 

Investigating the origin of these differences remains beyond the scope of this paper. We note that the two codes differ for the treatment of the physical processes underlying the resolution of the radiative transfer equation. The publicly available version of {\tt XSTAR}, for instance, does not include the treatment of continuum photoexcitation, an issue affecting, for example, the prediction of the Fe L lines in photoionized plasma (Guainazzi et al., in prep.). Previous benchmarks in the framework of AGN outflow studies \citep{mehdipour_2016} show differences up to 30\% in the optical depth of absorption lines produced at $\log \xi$ comprised between 1 and 2, and of about 20\% at higher ionization parameters.

\subsection{Future prospects with \textsl{XRISM} and \textsl{Athena}}
\label{sect:future}

\begin{figure}
\includegraphics[width=0.35\textwidth, angle=-90]{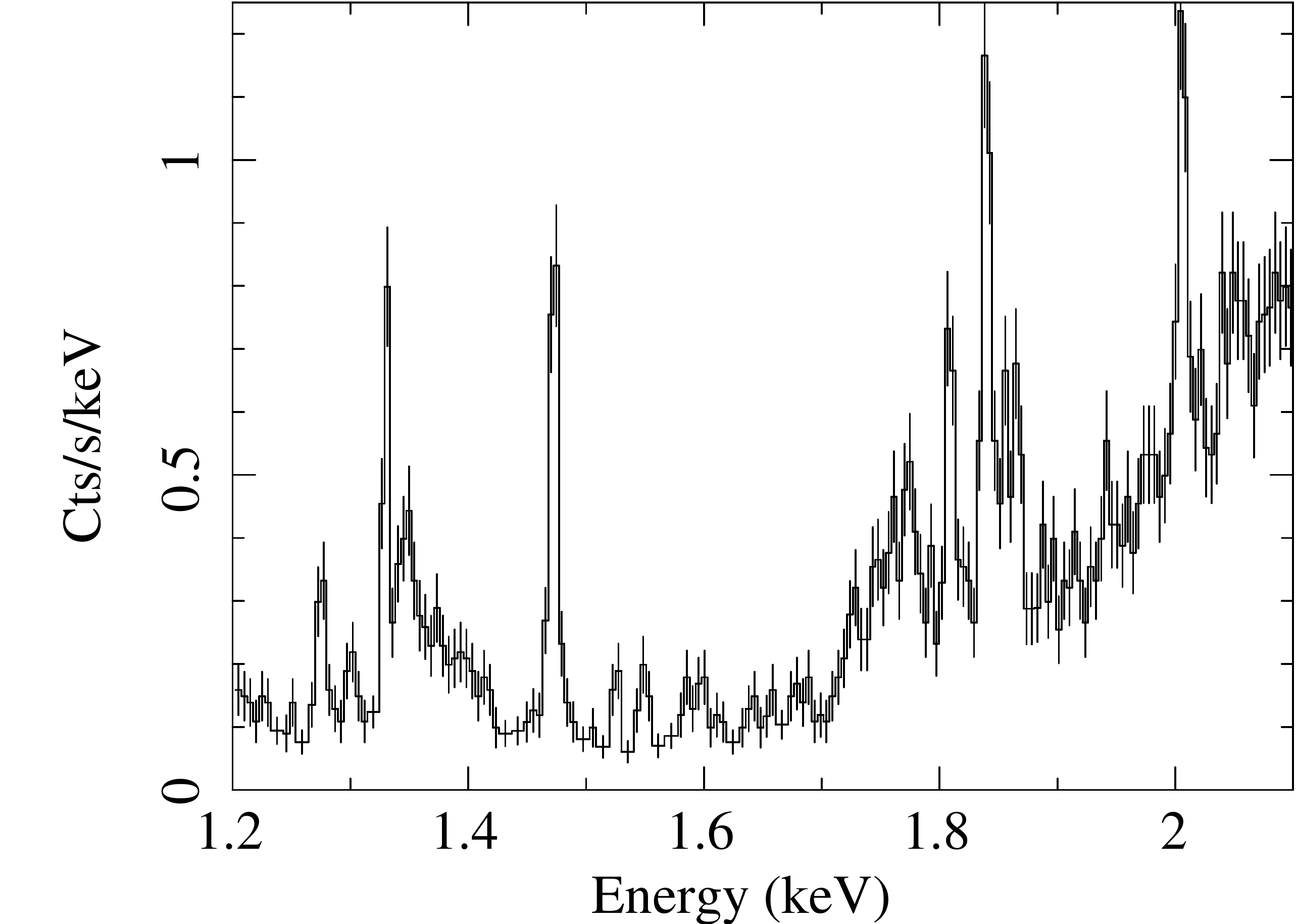}
\caption{Simulated spectrum of Vela~X-1 assuming the average pre-flare phase best-fit model presented in Tab.~\ref{tab:cloudy} and the instrumental response of the {\it Hitomi}/SXS instrument, the resolution of which was degraded to match the requirement of the XRISM/{\it Resolve}. Only the energy range between 1.2 and 2.1~keV is shown. The exposure time is 20~ks.}
\label{fig:xrism}
\end{figure}

The next decade will witness a transformational leap in the capability of high-resolution X-ray spectroscopy with the launch of the Japanese-led X-Ray Spectroscopy and Imaging Mission \citep[XRISM;][]{tashiro18} and of the ESA Advanced Telescope for High-Energy Astrophysics  \citep[\textsl{Athena};][]{nandra13} observatories. XRISM will carry a sensitive micro-calorimeter detector in the focal plane of a grazing incident telescope \citep[{\it Resolve}/SXS;][]{tashiro18} with an energy resolution of 7~eV (requirement) over the whole energy range between 0.2 and 12~keV. The {\it Resolve} instrument aims at bringing to full maturity the promises ushered by the ill-fated {\it Hitomi} mission, that was lost after six weeks of operations after yielding the first eV-class resolution spectrum of an extended celestial source \citep[the core of the Perseus Cluster;][]{hitomi16}. As a proxy of what {\it Resolve} spectroscopic observations could bring, we show in Fig.~\ref{fig:xrism}
the 1.2--2.1~keV 20~ks spectrum
that the {\it Hitomi/SXS} would have observed, assuming the best-fit model as in Tab.~\ref{tab:cloudy} and an energy resolution degraded to the {\it Resolve} requirements from the actual {\it Hitomi/SXS} resolution of $\sim$5\,eV that outperformed the instrument requirements \citep{Eckart_2018a}. The expected statistical accuracy of diagnostic ratios based on the He- and H-like lines of Mg and Si are: $\simeq$15\% for the Mg~{\sc xii}/Mg {\sc xi}, $\simeq$20\% for the Si~{\sc xiv}/Si~{\sc xiii}, $\simeq$25\% for the Mg~{\sc xi}~(r)/Mg~{\sc xi}~(f), and $\simeq$40\% for the Si~{\sc xiii}~(r)/Si~{\sc xiii}~(f) line ratios, respectively. 
This would correspond to a statistical error of 35\% and 45\% on the determination of the $R$ and $G$ parameters for Mg {\sc xi}, respectively. A 50~ks observation would improve the accuracy to 20\% and 30\%, respectively. The diagnostic parameters of the Si~{\sc xiii} line would be harder to constrain. One would need at least 30~ks to obtain a statistical accuracy better than 50\%. For a 50~ks observation, the estimated statistical accuracy for the Si~{\sc xiii} $R$ and $G$ parameters would be 25\% and 35\%, respectively.

The full power of phase-resolved spectroscopy will be unfolded with the launch of \textsl{Athena} \citep{nandra13}, thanks to the unprecedented combination of effective area and spectral resolution (2.5~eV at 7~keV) ensured by the X-Ray Integral Field Unit \citep[X-IFU; ][]{barret18}. 
\begin{figure}
\hspace{-0.5cm}
\includegraphics[width=0.35\textwidth, angle=-90]{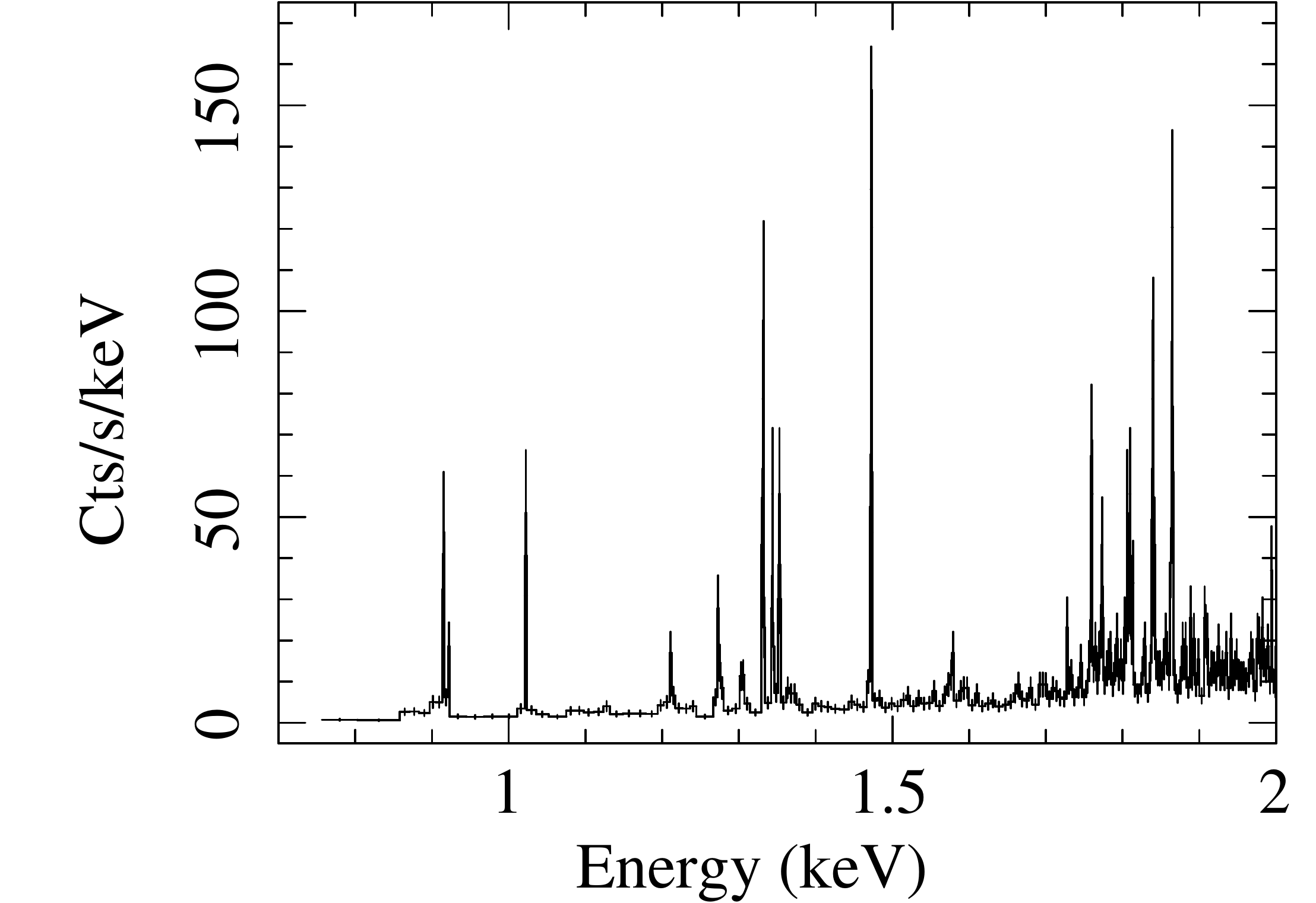}
\caption{Simulated \textsl{Athena}/X-IFU spectrum of Vela~X-1 assuming the average pre-flare phase best-fit model presented in Tab.~\ref{tab:cloudy}. Only the energy range between 0.8 and 2~keV, is shown. The exposure time is 300~s.}
\label{fig:athena}
\end{figure}   
In Fig.~\ref{fig:athena} we show a simulation of the predicted Vela~X-1 X-IFU spectrum corresponding to a net exposure time of 300~s (e.g., commensurate to the Vela~X-1 neutron star period). In the simulation, we  used the best-fit model as in Tab.~\ref{tab:cloudy}.  We assumed the current mirror configuration corresponding to effective area requirements of 1.4~m$^2$ at 1~keV, and 0.25~m$^2$ at 7~keV. Strong H- and He-like lines of Ne, Mg, and Si are visible by-eye, and their intensity and profile can be studied with an accuracy at the percent level. 

We note that both for \textit{Athena} and \textit{XRISM} simulations we used the nominal responses despite Vela~X-1 being a rather bright source. In the case of \textit{XRISM}, at the average observed flux level ($\sim$5\% of a Crab), the \textit{XRISM/Resolve} on-board processor should be able to process the data nominally. At the highest flux level ($\sim$10\% of a Crab), the processor could be marginally affected, but this is an unlikely occurrence. Regarding \textit{Athena}, the count rate of Vela~X-1 in the pre-flare state is $\sim$1900~counts/s, corresponding to about 20 times the counts expected from a mCrab source. At 50 mCrab, for instance, there is a penalty to pay in terms of throughput for the highest resolution events of $\sim$30\% \citep[see][for example]{peille_18}. This means that the nominal 300~s above correspond to effective 400~s exposure time.

\section{Summary \& outlook}
\label{sect:summary}

In the current study of the wind-accreting HMXB Vela X-1, we investigated the reaction of the clumpy stellar winds to an increase of the intrinsic emission from the neutron star.
The time-resolved analysis showed many emission and absorption features, especially prominent in the pre-flare phase against the low, highly absorbed continuum. Because of the strong increase of the continuum contribution during and after the flare, many of these line features disappear or weaken with regard to the continuum. 

In the spectra, we detect emission lines corresponding to H- and He-like ions of N, O, Ne, and Mg and Si. To our best knowledge, we present the first direct detection of O and N emission in Vela X-1.

The detected low ionization of Mg K$\alpha$ absorption lines indicate  the presence of colder, less ionized material in the line of sight potentially originating from the clumpy winds or parts of the accretion structure, although the latter is not expected to be strong at the given orbital phase.

The RRC feature of oxygen yielded a plasma temperature of $T_\textnormal{e}$ $\sim$1.5$\times 10^5$~K, which is typical for photoionized plasmas.

Our attempt to probe the density and temperature of the wind plasma through the $R$ and $G$ diagnostic parameters resulted in large uncertainty ranges, especially in the flare and post-flare phases, while the wind velocity estimations appeared to be consistent with zero. 

The best-fit model with two \texttt{CLOUDY} components calculated for a photoionized plasma suggests the presence of two distinct gas regions with high and low ionization levels. A strong UV emission from the supergiant companion is required to achieve a good agreement between models and data. Spectral modeling with \texttt{XSTAR} delivers the best result in combination with a collisionally ionized  \texttt{XSPEC} model \texttt{vapec}.

By simulating Vela X-1 spectra as it might be seen by \textit{XRISM} and \textit{Athena}, we show that this source is an ideal target for observations by the future missions that have the potential to probe plasma properties with a much higher accuracy and precision.

\begin{acknowledgements}  We would like to thank the anonymous referee for providing useful feedback. VG is supported through the Margarete von Wrangell fellowship by the ESF and the Ministry of Science, Research and the Arts Baden-W\"urttemberg. We acknowledge support from the ESTEC Faculty Visiting Scientist Programme to VG. SB acknowledges financial support from the Italian Space Agency under grant ASI-INAF 2017-14-H.O. Work at LLNL was performed under the auspieces of the U.S. Department of Energy under contract No. DE-AC52-07NA27344. SMN acknowledges funding by the Spanish Ministry MCIU under project RTI2018-096686-B-C21 (MCIU/AEI/FEDER, UE), co-funded by FEDER funds and by the Unidad de Excelencia María de Maeztu, ref. MDM-2017-0765. This study made use of following software and packages:  \texttt{XSPEC} \citep{arnaud_96}, \texttt{PyXspec}, \texttt{astropy}\footnote{http://www.astropy.org} a community-developed core Python package for Astronomy \citep{astropy:2013, astropy:2018}, \texttt{matplotlib} \citep{matplotlib},  \texttt{numpy} \citep{numpy}, \textit{XMM-Newton} SAS \citep{gabriel_04}, and NASA's Astrophysics Data System Bibliographic Services.
\end{acknowledgements}

\bibliographystyle{aa}
\bibliography{aa_abbrv,mnemonic,references}

\end{document}